# 3D photophoretic aircraft made from ultralight porous materials can carry kg-scale payloads in the mesosphere

Thomas Celenza*, Andy Eskenazi* and Igor Bargatin

*Department of Mechanical Engineering and Applied Mechanics, University of Pennsylvania, Philadelphia, Pennsylvania, USA*

*We show theoretically that photophoretic aircraft would greatly benefit from a three-dimensional (3D) hollow geometry that pumps ambient air through sidewalls to create a high-speed jet. To identify optimal geometries, we developed a theoretical expression for the lift force based on both Stokes (low-Re) and momentum (high-Re) theory and validated it using finite-element fluid-dynamics simulations. We then systematically varied geometric parameters, including Knudsen pump porosity, to minimize the operating altitude or maximize the payload. Assuming that large vehicles can be made from previously demonstrated nanocardboard material, the minimum altitude such vehicles can levitate at is approximately 55 km, while the payload can reach approximately 1 kilogram at 80 km altitude for vehicles with 10-meter diameter. In all cases, the maximum areal density of the sidewalls cannot exceed a few grams per square meter, demonstrating the need for ultralight porous materials.*

For centuries, human exploration of Earth's atmosphere and outer space has led to important advancements in fields such as aerodynamics, astronomy, and climate modeling [1-3]. However, some atmospheric regions remain less understood due to limitations in existing propulsion technologies. One such area is the mesosphere, which lies between 50 to 80 kilometers above Earth. In this layer, rising levels of carbon dioxide are paradoxically causing rapid cooling [4]. This cooling, in turn, causes the atmosphere to contract, resulting in reduced satellite drag and increased space debris [5, 6]. The challenge of studying these effects lies in the mesosphere's unique conditions: its air pressure is too low for planes or balloons but too high for orbiting satellites. As a result, there are large uncertainties in our understanding of the mesosphere and related phenomena [7].

The Martian atmosphere is another area of keen interest, as recently highlighted by the near-surface flight of the Ingenuity helicopter [8]. However, achieving sustained flight at higher Martian altitudes (> 10 km) remains challenging due to the extremely low atmospheric density [9,10]. Similar to the obstacles faced in Earth's mesosphere, the study of Mars' higher elevations is limited by the lack of long-duration propulsion systems that can operate under very low-pressure conditions (less than 1 mbar or 100 Pa). Developing an airborne platform capable of functioning in such thin atmospheres, on both Mars and Earth, would be invaluable for collecting essential data on wind patterns, temperature and pressure fluctuations, and atmospheric gas concentrations.

Light-powered lightweight levitating plates developed by Cortes *et al.* [11] can potentially achieve sustained flight in both Earth's mesosphere and the Martian atmosphere. These microflyers are made from "nanocardboard," extremely lightweight porous plates, and can levitate using photophoresis, a light-driven propulsion method that leverages Knudsen pumping to create a jet of ambient gas. The phenomenon is strongest when the mean free path of molecules is comparable with the characteristic dimension of the channels the gas flows through. Knudsen pumping pushes air through elongated channels or capillaries upon exposure to incident radiation, such as light, because the absorption of light induces a temperature gradient along the channel's length. Air molecules coming from the warmer side of the channel transfer more momentum to the channel when they hit the channel sidewall (**Fig. 1f**). Consequently, a net shear stress is produced on the channel sidewall towards the cooler side, generating a corresponding reaction force on the gas directed towards the warmer side. This results in the migration of molecules from cold to hot, even against a small opposing pressure gradient, which is the essence of pumping [12,13]. Knudsen pumps operate without moving parts, relying on temperature gradients to drive gas flow through the porous plates.

Photophoretic levitation typically relies on a difference in physical properties between a plate's top and bottom. In the research conducted by Cortes *et al.* [11], the bottom side of the nanocardboard was coated with carbon nanotubes (CNTs), which absorbed the incoming light and became hotter than the top side. The nanocardboard material consists of ultra-thin aluminum oxide face sheets, ranging from 25 to 400 nm in thickness, connected by micro-channels. This structure gives them an extremely low areal density of approximately 1 g/m² and a bending stiffness several orders of magnitude higher than solid plates of the same mass [14]. Under illumination, the temperature gradient triggered Knudsen pumping, forcing air to move from the cooler top through the nanocardboard channels to the warmer bottom. The exiting air created a downward jet below the nanocardboard plate, producing a reaction lift force capable of levitating centimeter-scale sized plates and carrying tiny "smart dust" sensor payloads [11,15]. This propulsion mechanism is most effective in low-pressure environments (1-100 Pa) [16], making it suitable for applications in Earth's mesosphere and high-altitude regions on Mars, such as Olympus Mons [17]. Swarms of these microflyers could be deployed on Earth or Mars to gather important upper atmosphere data.

---

* Both authors contributed equally to this work.

In this work, we propose large three-dimensional photophoretic vehicles with several meters wide diameters and porous sidewalls that channel air into a central chamber and expel it through a bottom nozzle (**Fig. 1**). Using the nozzle increases the speed of the exiting air jet, which in turn results in a higher lift force and widening of the range of operating pressures where flight can be achieved. The larger size and higher jet speeds means that such jets can operate over a range of Reynolds numbers from deep in the Stokes regime ($Re \ll 1$) to the momentum theory regime ($Re \gg 1$), which requires new modeling approaches.

The use of Knudsen pumping to create high-speed jets is a novel physics principle for propulsion. Photophoresis has previously shown promise at the microscale and cm-scale [11,16], but has never been considered at the meter scale like we are proposing in this work. We draw on our previous theoretical models on levitating planar nanocardboard [11] and solid mylar-CNT composite disks [18] to analyze these 3D vehicles below.

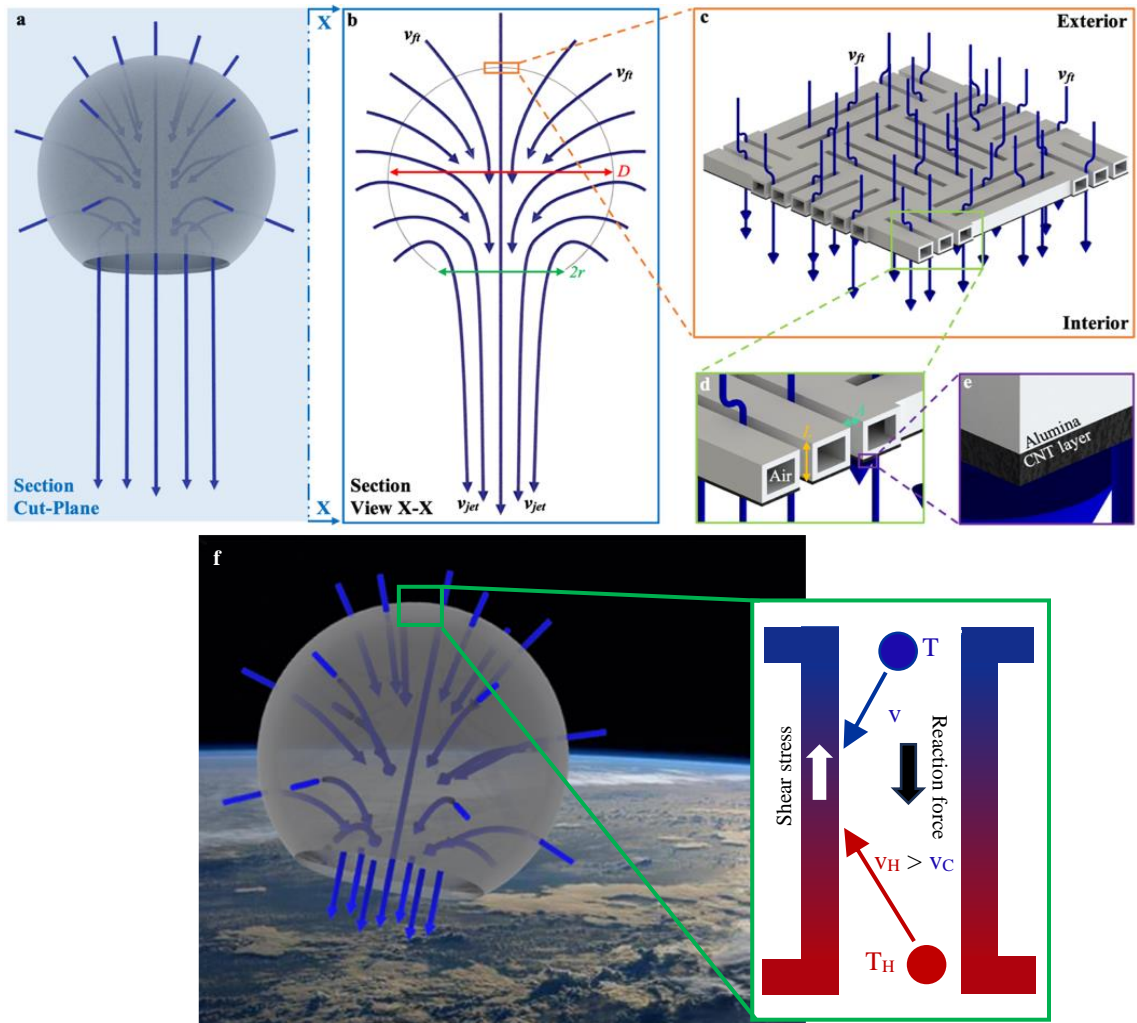

*Figure 1*: *A hollow sphere with porous alumina-CNT composite walls (a). The cross-sectional view (b) illustrates air being drawn into the sphere through its porous walls due to Knudsen pumping and then expelled as a jet through the exit nozzle. The velocity with which the air flows through the porous walls is labeled $v_{ft}$, as detailed in the zoomed-in view (c), while the exiting jet velocity is labeled $v_{jet}$. The dimensions are defined as follows: A is the nanocardboard channel width, L the nanocardboard channel height, and r the structure's outlet radius, while D is the structure's overall size dimension. As (d) demonstrates, the alumina walls are notably thin, with 50-nm wall thickness, about one order of magnitude smaller than the channel width and length. The final view (e) highlights a thin layer of CNT on the*

*sphere's interior, which absorbs light but has minimal impact on air flow. Panel (f) shows the molecular effects occurring within the channel to induce Knudsen pumping, alongside an example model of flight in the mesosphere.*

The walls of our proposed 3D vehicles are made of porous nanocardboard, consisting of alumina and coated with carbon nanotubes (CNTs) on the inner side. Because alumina is transparent, only the CNTs absorb incoming light, inducing Knudsen pumping of air from the outside into the interior chamber through the channels and out of the chamber through the exit nozzle, producing a jet as illustrated in **Fig. 1**. The buildup of air inside the structure creates a small overpressure, which is accelerates the air through the exit nozzle but does not significantly slow down the Knudsen pumping through the walls of the structure, as shown by our numerical models detailed in the Supplementary Information.

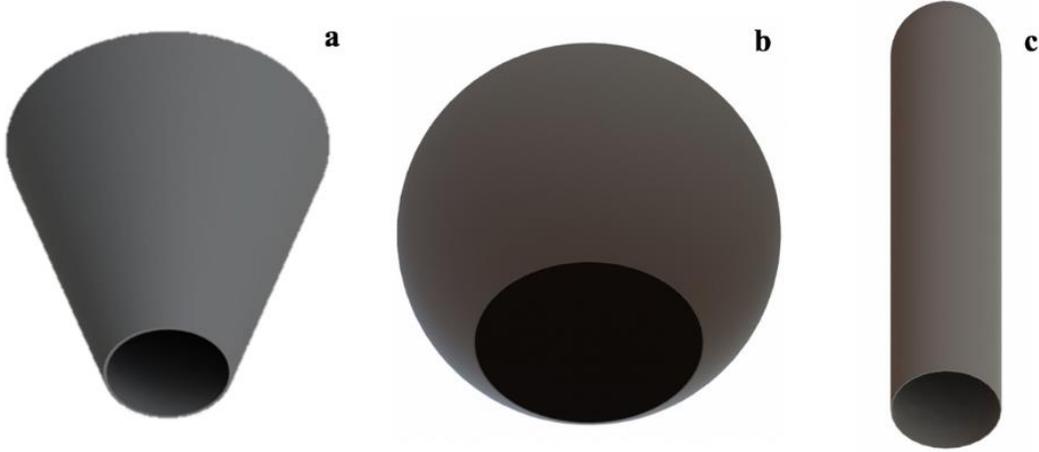

*Figure 2: The three representative shapes explored in this work, a cone (a), a sphere (b) and a rocket (c). Here, the nozzle view for all three geometries is shown.*

To identify the optimal 3D geometry that maximized payload, we considered three representative shapes (a sphere, a cone, and a rocket, shown in **Fig. 2**), and conducted simulations to determine the parameters that would yield the greatest lift forces. First, we needed to find an analytical expression to estimate reactive lift forces from the jet exiting the nozzle over a broad range of Reynolds numbers. We used computational fluid dynamics simulations in ANSYS Fluent to model the lift produced by these 3D vehicles, considering outlet jet velocities ranging from $10^{-6}$ m/s to $\sim 10^2$ m/s and at atmospheric altitudes from 0 to 80 km. This allowed us to obtain jet reaction lift forces for Reynolds numbers across nine orders of magnitude, from deep in the Stokes regime, at $Re \sim 10^{-4}$ for the smallest structures and flow velocities, to $Re \sim 10^4$ for the largest structures and flow velocities.

The simulations were axisymmetric, rotating a 2D model around a central axis to generate 3D results. They were conducted in an air box 10 to 100 times larger than the vehicle itself to ensure accurate lift force calculations without interference from the walls. Additional details can be found in the supplementary information, but **Fig. 3** provides an overview of the modeling environment and boundary conditions. The simulation employed ANSYS Fluent's SST k-omega solver with standard settings, except for enabling low-Reynolds number corrections. The lift force was found by integrating the forces acting on the boundary of the airbox, which in the steady state, were equal and opposite to the forces created by the simulated 3D vehicle.

In the ANSYS simulations, the inner wall (colored red in **Fig. 3a**) acted as the inlet of the simulated volume, drawing air into the central chamber at a given flow-through velocity, $v_{ft}$. The outer wall (violet) served as the outlet, with the air flow exiting the simulated volume at the same flow-through velocity $v_{ft}$. This setup allowed us to estimate the lift force without the computational burden of simulating the air flow through the structure's microchannels. Instead, we replaced the flow through the microchannels with uniform flow at an effective flow velocity $v_{ft}$ through the porous wall. Effectively, in the ANSYS simulations, we treated $v_{ft}$ as an independent variable to find the lift force as a function of $v_{ft}$ for different shapes of the vehicle (the actual values of $v_{ft}$ for each vehicle and altitude were found later using a MATLAB model of Knudsen pumping described below). These simulations produced a dataset of reaction forces across a wide range of operating altitudes, flowthrough velocities $v_{ft}$, and jet velocities $v_{jet}$.

Next, we fitted the collected data to the following equation:

$$F = C_1 8\mu D v_{ft} + C_2 \rho K v_{jet}^2, \tag{1}$$

where $\mu$ represents the fluid viscosity, $\rho$ the atmospheric air density, $K = \pi r^2$ defines the area of a nozzle with radius $r$, $D$ is the characteristic (i.e., largest) dimension of the geometry, $v_{ft}$ the flow-through velocity through the porous walls, and $v_{jet}$ is the velocity of the fluid exiting the structure through the small nozzle. As detailed in the supplementary information, $v_{ft}$ depends on the light intensity, $I$, the altitude-dependent air pressure, $P$, the geometric parameters of the nanocardboard and the vehicle's 3D shape. The upper limit of the flow-through velocity scales as $v_{ft} \approx 0.03\, I/P$ (see supplementary information), resulting in velocities of less than 1 mm/s under natural sunlight (~1000 W/m$^2$) and standard atmospheric pressure ($10^5$ Pa) but this speed can increase by many orders of magnitude as the pressure drops at higher altitudes. All altitude-dependent atmospheric properties used in simulations and numerical calculations were based on Ref. [19].

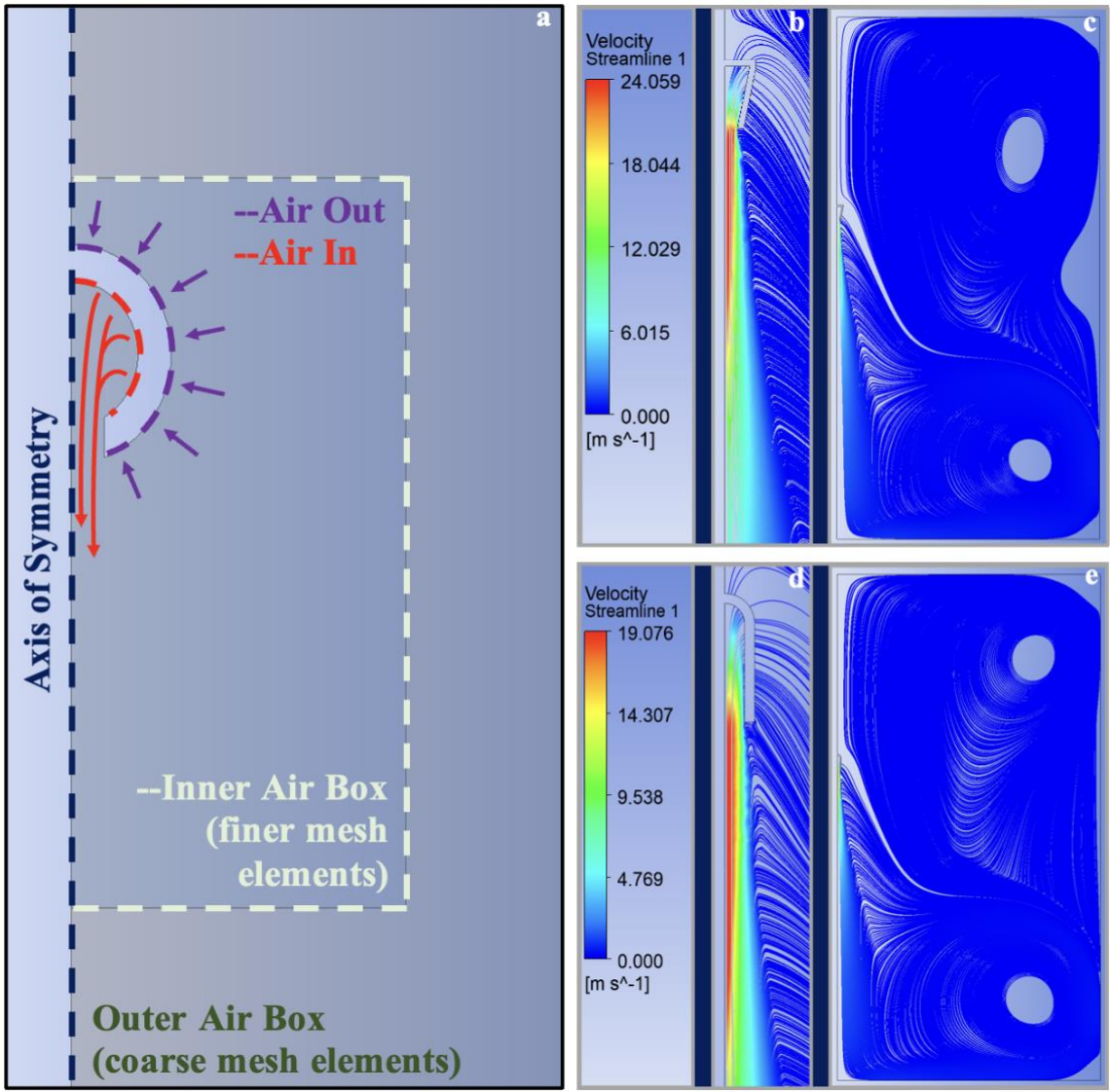

*Figure 3*: *(a) Boundary conditions implemented on the ANSYS simulations, using a sphere as an example. The inner air box region (whose boundary is shown in light green) was employed for finer meshing close to the vehicle. The outer air box, not fully shown in the figure, extended to at least 10 times the size of the sphere. This size was generally adequate for the lift force to stabilize, providing a reliable estimate for the expected mid-air lift force. (b)-(e) show the velocity streamlines corresponding to the cone (b, c) and rocket (d, e) geometries simulations in ANSYS, for a flow-through velocity of 1 m/s and atmospheric conditions corresponding to 0 km in altitude. Both the cone and rocket have a characteristic dimension (D) of 5 cm. (c) and (e) show the entire computational domain, i.e., a zoomed-out view of plots (b) and (d), respectively.*

The first term in Eqn. (1), which is dominant at low *Re*, is based on prior work of Cortes et al. [11] (see also SI, page 14), who showed analytically that the lift force for nanocardboard disks that pump gas through them at low *Re* was equivalent to the drag experienced by a solid disk moving through a stationary fluid at the same velocity $v_{ft}$. Briefly, the argument is as follows: If we switch to the frame of reference moving vertically down at $v_{ft}$, the average velocity of the gas near the surface of the object becomes approximately zero. Therefore, the boundary conditions are the same as for a solid plate without any channels, though the boundary is not stationary in this moving frame and instead moves upward at the flow-through velocity. In the Stokes flow approximation, there is no explicit time dependence. Hence, the solutions for a moving boundary condition are the same as for a stationary boundary condition and we can use the known drag equations for a disk and other shapes in the Stokes regime [20], all of which have the form matching the first term in Eqn. (1).

The second term in Eqn. (1), dominant at high *Re*, is rooted in the helicopter-momentum theory equation, which is an application of Reynolds Transport Theorem. Since the viscosity forces become less important at high speeds, this term simply represents the momentum transferred to the exiting jet. Summing both terms results in a simple interpolation between the low-Re and high-Re operating regimes, providing an estimate for the lift force at all pressures and velocities. The constants $C_1$ and $C_2$ in Eqn. (1) were fitted to match the simulated lift forces for various shapes we studied but, as detailed in the supplementary information, the fitted values $C_1$ and $C_2$ were always on the order of 1. **Table 1** summarizes the average fitted $C_1$ and $C_2$ parameters obtained from fitting the results for 27 ANSYS Fluent simulations using 3 different altitudes (0 km, 40 km and 70 km), 3 geometry types (sphere, cone, and rocket), and 3 different structure sizes (1cm, 5cm and 10cm).

| Fitting Parameters for Each Geometry | | | |
|---|---|---|---|
| **Geometry** | **Cone** | **Sphere** | **Rocket** |
| $C_1$ | 1.2 | 1.3 | 1.4 |
| $C_2$ | 0.9 | 0.9 | 0.4 |

*Table 1: Fitting parameters for the three simulated geometries.*

The lift Eqn. (1) uses only continuum regime terms because, for the most promising vehicle shapes, the mean free path of the gas molecules is orders of magnitude smaller than the dimensions of the vehicle itself. For example, the mean free path at 80 km is on the order of 1 cm (and smaller at lower altitudes), while the optimal structures we consider are approximately 10 meters in size. We note that we do not model the microchannels in the ANSYS simulations. Instead, the analysis of rarefied gas flow through microchannel was completed in the MATLAB numerical analysis that is valid in both continuum and free molecular regimes. We also find that, even at operating altitudes in the mesosphere, this pressure is small compared to the ambient pressure (> 1 Pa), allowing us to use incompressible flow assumption for the analysis of airflow.

After establishing that Eqn. (1) with $C_1 = C_2 = 1$ as a reasonable estimate of the lift force (accuracy of 20-30%), we focused on fine-tuning various parameters of the overall vehicle shape and the porous microstructure of the nanocardboard to enhance its payload capacity. The developed MATLAB code [21] was based on the photophoretic levitation theory originally developed for nanocardboard [11] and adapted to axisymmetric 3D structures, as detailed in the supplementary information. The code also accounted for the altitude-dependent variations in temperature and pressure based on standard atmospheric models. Our optimization aimed to find the best set of parameters, including *A* (width of the nanocardboard channel), *L* (height of the nanocardboard channel), and *r* (radius of the structure's outlet or nozzle), that would either maximize payload or enable flight at the lowest possible altitude. This maximum payload and minimum flight altitude were evaluated as a function the overall size of the aircraft, represented by *D* (the diameter for a sphere and cone, or the length for a rocket, as shown in **Fig. 5**). To be able to use the formulas for rarefied gas flow [11], we also enforced the constraint *L* > *A*, i.e., nanocardboard channels need to be longer/taller than they are wide.

Our computational analyses showed that the best values for the nanocardboard porosity parameters *A* and *L* remained fairly consistent across all shapes and sizes. Specifically, when the mission goal was to achieve flight at the lowest possible altitude (55 km without carrying any payload), the optimal values for *A* and *L* were approximately 0.20 mm and 0.21 mm. When optimized for maximum payload (achieved at 80 km altitude), *A* and *L* increased to 0.90 mm and 0.91 mm, or about a factor of 4.5 greater. Despite a roughly 40-fold change in ambient pressure between these two altitudes (55 km and 80 km), the optimal *A* and *L* values were of the same order of magnitude (sub-millimeter). This result suggests that it is feasible to design structures capable of levitating in the lower mesosphere while still being able to carry a significant payload at higher altitudes.

The maximum areal densities (i.e., vehicle mass divided by total nanocardboard sidewall area) that could be levitated were also comparable for all structures. **Table 2** shows that the typical value of maximum areal density was ≈ 7.1 g/m$^2$ (grams per square meter) for smaller aircraft with a diameter of $D$ = 10 cm and ≈ 5.5 g/m$^2$ for larger aircraft with a diameter of $D$ = 10 m. **Fig. 4a** illustrates how these maximum areal densities changed with the size of the aircraft ($D$) and the airflow's Reynolds number ($Re$). The permissible areal densities of each structure decreased with increasing size and $Re$ and stabilized at ~5.5 g/m$^2$ for larger aircraft capable of carrying payloads of 1 gram or more. These maximum areal densities were similar to the order-of-magnitude estimate of 4 g/m$^2$, which was derived in the supplementary information. This estimate was calculated as $C_3 I/(v_{avg} g) \approx 0.004$ kg/m$^2$, where $I$ = 1000 W/m$^2$ is the incident optical intensity, $v_{avg} = \sqrt{8 R_{air} T / \pi} \approx 400$ m/s is average speed of air molecules at 55-80 km altitudes, $C_3 \approx 0.016$ is a constant, and $R_{air} = R_u / M_{air} = 287$ $J/(kg \cdot K)$ is the gas-specific ideal constant of air, equal to the universal gas constant $R_u$ divided by the average molar mass of air $M_{air}$.

| | | Areal Densities and Areas Ratio | | | | | |
|---|---|---|---|---|---|---|---|
| **Geometry** | | **Cone** | | **Sphere** | | **Rocket** | |
| | | **D = 10 cm** | **D = 10 m** | **D = 10 cm** | **D = 10 m** | **D = 10 cm** | **D = 10 m** |
| **Max Areal Density** | For Max. Payload | 6.6 g/m$^2$ | 5.4 g/m$^2$ | 7.8 g/m$^2$ | 5.5 g/m$^2$ | 6.9 g/m$^2$ | 5.7 g/m$^2$ |
| **Area Ratios** | For Min. Altitude | 18 | 26 | 26 | 27 | 23 | 25 |
| | For Max. Payload | 5 | 5 | 5 | 6 | 6 | 6 |

*Table 2: Summary of the parametric studies results for the Cone, Sphere and Rocket geometries, for values of D = 10 cm and D = 10 m (full data for all values of D can be found in the supplementary information section). Here, the area ratio refers to the $K_{total}/K_{out}$ ratio, of the structure's total surface area to its outlet area.*

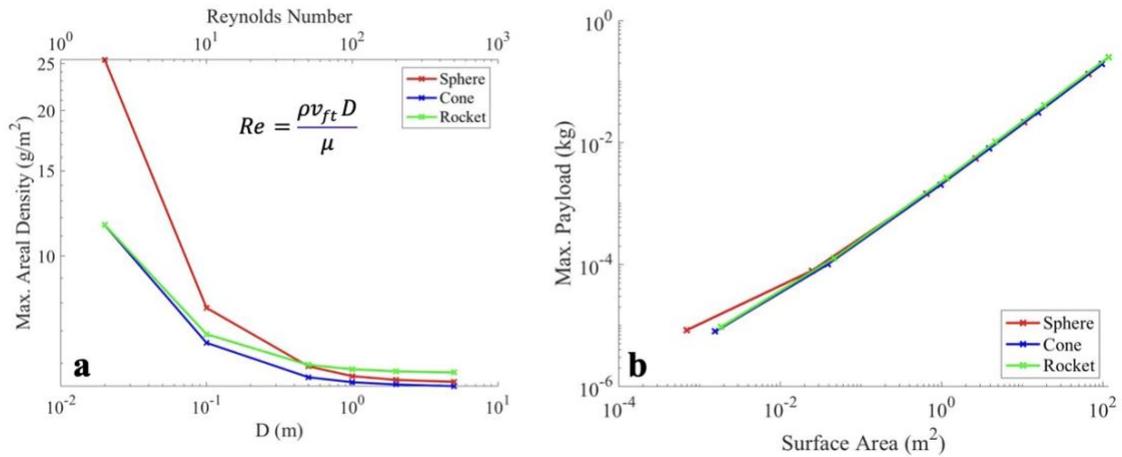

*Figure 4: (a) Maximum areal density versus characteristic size and (b) maximum payload versus surface area for the three considered 3D geometries at 80-km altitude. Each data point corresponds to the optimized geometry at each of the probed values of the parameter D. The overlap between the curves, in particular at surface areas larger than 0.01 m$^2$, shows that all three geometries have similar areal densities and maximum practical payload capabilities.*

Assuming practical nanocardboard has an areal density of 1 gram per square meter [11], we plotted the maximum payload against the structure surface area in **Fig. 4b**. While the sphere performed best at smaller sizes, all three shapes delivered nearly identical performance at larger sizes, which can carry the largest payloads and have the most practical applications. While the structure could be folded to increase the total area, parts of the structure will be shaded or experience very low sunlight intensity due to oblique incidence of sunlight. As a result, we chose shapes that do not self-shadow, are easiest to fabricate and are most likely to maintain their shape during the flight. For example, the sphere and cylinder, and to a smaller extent the cone, are all robust with respect to deformations due to the overpressure from Knudsen pumping.

**Fig. 5** below illustrates optimized shapes for the 10-meter cone (a), sphere (b) and rocket (c), which could carry 780, 540, and 1020 grams of payload, respectively, sufficient for modern communication devices [22] or even typical CubeSats [23]. As highlighted in **Table 2** and the supplementary information section, we observed that the ratio of the total surface to the outlet area, $K_{total}/K_{out}$, was also fairly consistent for the optimal geometries. In the case of achieving flight at the minimum altitude of 55 km, this ratio varied between 17 and 42, with an average value of about 23 across all shapes and sizes. For scenarios aiming for maximum payload at an altitude of 80 km, the typical ratio was around 6, as depicted in **Fig. 5**.

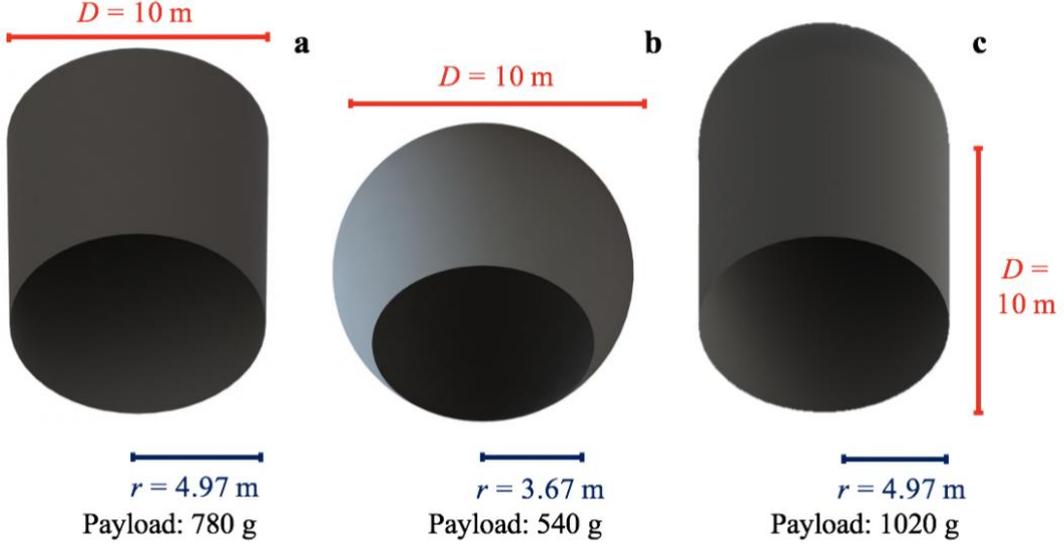

***Figure 5***: *(a) Cone, (b) sphere, and (c) rocket vehicles geometrically optimized for maximum payload capabilities for a characteristic dimension of D = 10 meters. D refers to the diameter of the cone and sphere, and the length of the rocket. To achieve a 1 kg payload, the cone and sphere required diameter D of 11.5 meters and 14 meters, respectively. The optimal angle for the cone, as seen in panel (a), is quite small, making it appear more cylindrical in shape.*

Due to the principle of mass conservation and the incompressible fluid assumption, the speed of the jet exiting the outlet must be greater than the velocity of air flowing through the channels by the $K_{total}/K_{out}$ area ratio. Therefore, recalling the $v_{ft} \approx 0.03\ I/P$ relationship, we can approximate $v_{jet} = v_{ft} K_{total}/K_{out} \approx 0.18\ I/P \approx 0.18 \times 1300\ W\ m^{-2}/1\ Pa = 234\ m/s$ at the maximum payload altitude of 80 km, while at the minimum altitude of 55 km (for zero payload), $v_{jet} = v_{ft} K_{total}/K_{out} \approx 0.70\ I/P \approx 0.70 \times 1200\ W\ m^{-2}/10\ Pa = 84\ m/s$. We note that for the payload altitude of 80 km, the jet speed approaches but remains below the speed of sound, given by $v_{sound} = \sqrt{\gamma R_{air} T_{80km}} \approx \sqrt{1.4 \times 287\ J/(kg\ K) \times 200\ K} \approx 280$ m/s, where $\gamma$ is the adiabatic constant of air and $T_{80km} \approx 200\ K$ is the air temperature at 80 km altitude.

Achieving kg-scale payloads in the mesosphere therefore requires building 10m-scale photophoretic aircraft out of ultralight materials that simultaneously possess low areal densities ($\approx 1$ g/m$^2$) and sufficient structural integrity. Nanocardboard has areal densities of approximately 1 g/m$^2$ as well as bending stiffness values 4 orders of magnitude higher than those of flat plates with the same mass. In fact, because its bending stiffness improves with channel length in the same way as in sandwich composite plates, nanocardboard can achieve bending stiffness values of up to 3×10$^{-6}$ N·m and, in addition, recover its shape after sharp bending [14]. These shape recovery properties of nanocardboard can help a large photophoretic aircraft recover from buckling and other deformations caused by winds or non-uniform radiation. In the Supplementary Information, we included COMSOL simulations of a 10-meter nanocardboard sphere slightly deforming but not buckling under its own weight. In contrast, the other geometries are predicted to buckle under their own weight, which indicates the need for additional lightweight support structures or frames which could be made of carbon fiber or similar lightweight, but strong materials.

These proposed aircraft structures do not necessarily have to be rigid; flexible designs using parachute or balloon-like structures could also be effective, as illustrated by the balloon-like shapes in **Fig. 5**. For such designs, porous Mylar material may be a viable option. Nanocardboard-like panels of Mylar could meet the structure's requirements, and Mylar is already used in many NASA balloon missions [24,25]. However, alumina is superior to Mylar in terms of resistance to UV radiation and mechanical stiffness. All calculations in this work therefore assume the material is alumina, just like nanocardboard. In order to maintain an upright vertical axis, the carried payload would be suspended from the bottom of the structure with a tether, lowering the center of gravity, and maintaining orientation during flight.

We made several key assumptions about the pressure and density inside the structure. First, we neglected the pressure buildup inside the structure, which is caused by airflow through the channels. Second, we made the simplifying assumption that air was incompressible when calculating the jet velocity. **Fig. 6** shows the pressure differential between the inside and outside of the balloon, showing the change is minimal. Even at operating altitudes in the mesosphere, this pressure is small compared to the ambient pressure (> 1 Pa), also allowing us to use incompressible flow assumption for the analysis of airflow. The largest difference in air pressures is ~20% at the highest altitudes, around 80 km. While this pressure is sufficient to drive airflow out of the exit nozzle for different geometries, it is a small differential that does not alter the air properties enough to impact our calculations since our lift estimate Eqn. (1) already involves errors of 20-30%..

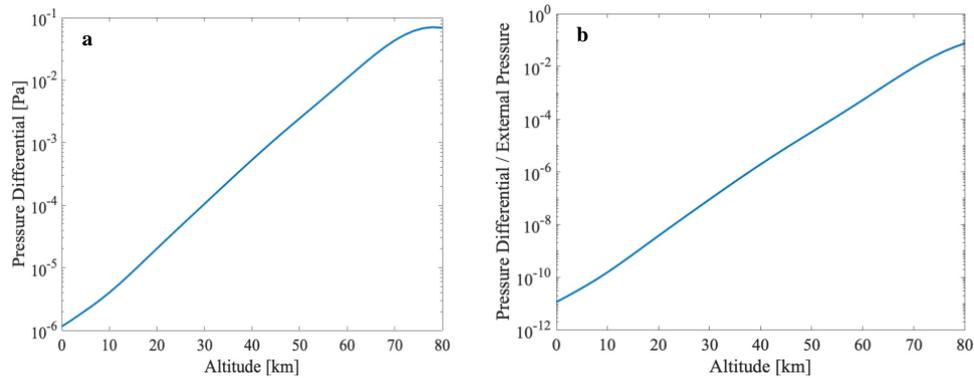

*Figure 6*: *This pressure difference and pressure difference divided by external pressure represents the variance between the pressures inside and outside of the 3D structure, which is also the pressure forcing flow out of the outlet nozzle. The structure chosen for the above plot is the optimal spherical structure for maximum payload from earlier in the paper.*

In our calculations, we assumed a light intensity of 1000 W/m$^2$ illuminating all surfaces. For reference, the direct sunlight intensity in the mesosphere is approximately 1360 W/m$^2$, similar to outer space. Additionally, Earth's planetary albedo of ~0.3 contributes an extra ~500 W/m$^2$ of reflected sunlight from clouds and the Earth's surface below the aircraft. Depending on the Sun's elevation and the surface orientation, the aircraft could be exposed to a range of light intensities, from zero to nearly 2000 W/m$^2$, combining both direct and reflected sunlight. If the aircraft rotates due to atmospheric winds, as balloons usually do, all surfaces will experience an average light flux ~1000 W/m$^2$. Non-uniform radiation from the sun causes be a variation in the rate of Knudsen pumping through the walls of the 3D photophoretic vehicle, which reduces the jet velocity and the payload compared to uniform illumination assumption but does not disable the propulsion mechanism we propose. As a point of reference, we also conducted simulations with a reduced light intensity of 500 W/m$^2$. In these scenarios, the achievable payloads were about four times lower than those calculated under the 1000 W/m$^2$ assumption.

Additionally, although higher pressures increase the density of air inside the vehicle, their effect is counterbalanced by the air inside the vehicle being warmer, resulting in a density difference of less than 20%. At certain altitudes, the air inside the vehicle may even become less dense than the surrounding air, creating a buoyancy effect at altitudes below 50 km, i.e., below the mesosphere. In the mesosphere, the air density is so low that buoyancy effects have little-to-no impact on the structures we are examining. We also note that we conservatively assumed nanocardboard thickness and channel heights that we previously fabricated [11]. If future innovations enable the creation of even thinner, porous films, it will allow the photophoretic 3D flyers to operate at even lower minimum altitudes than 55 km, closer to the 35-40 km range.

An important constraint for photophoretic aircraft is that the need for continuous light exposure to generate lift. This condition limits their operational window to approximately 12 hours a day in low and mid latitudes, after which they begin to descend. However, near the poles, extended periods of daylight during the polar day allow for operations lasting weeks or months. Despite these limitations, these aircraft are valuable for gathering data in the mesosphere because current technologies, like sounding rockets, can only remain airborne in this region for a few minutes [8,10,26]. Once the mission concludes, payloads can be safely returned to Earth, similar to how radiosonde weather balloon payloads descend using a parachute. Given their shape, the 3D vehicles depicted in **Fig. 5** could serve as parachutes for this purpose.

For deploying our proposed photophoretic vehicles, research balloons currently offer the most practical method. These helium balloons can ascend to roughly 40 km [26], just below the minimum altitude where photophoretic aircraft are effective with the 50 nm alumina thickness we assumed. After reaching peak altitude, the helium balloon can release the aircraft, at which point the photophoretic Knudsen pumping takes over as the main lifting force, allowing the aircraft to rise to their ideal sensing altitudes. Another potential option is to use sounding or suborbital tourism rockets, which can reach altitudes of 100 km or more [27]. These rockets can deploy payloads at lesser altitudes, making them an apt choice for releasing photophoretic aircraft as long as they can move sufficiently slowly to release the vehicle gently, avoiding any damage or tears to the structure.

High-resolution measurements of winds, temperature, pressure, magnetic fields, and gas concentrations are some of the most valuable data that could be gathered in the mesosphere [28-31]. A network of these flying vehicles could provide unparalleled temporal and spatial resolution in monitoring these parameters, thereby enhancing our capabilities in climate and weather modeling. The payloads we have considered could be equipped with lightweight Iridium transmitters for communication and either batteries or photovoltaics for power, allowing for real-time data transmission back to Earth. Alternatively, the data could be stored and the payload retrieved once the aircraft descends. This option could be particularly useful in situations where equipment reuse is desirable.

The main focus of this paper is on the theoretical modeling of a novel propulsion method based on Knudsen pumping. While we are in the process of conducting experiments to validate the theoretical model discussed, the fabrication of large structures—measuring 10 meters or more—is currently impractical given our available resources. This is particularly challenging because the required materials would need to have microscale features spread across a surface area of hundreds of square meters. However, we anticipate being able to test smaller structures, on the scale of centimeters, and some initial results were presented in Ref. [32].

In summary, we demonstrate that spherical photophoretic aircraft constructed with ultralight, ultrathin, porous materials have the capability to carry payloads on the scale of kilograms, which is similar to the mass of CubeSats, without the need for moving parts or fuel. Our work introduces new applications of the photophoretic force for 3D structures to enhance propulsion and presents a mathematical model that explores varying geometries through innovative theoretical approaches. The findings from this study can also be extended to high-altitude operations on Mars by adapting a Martian atmospheric model [33]. This research paves the way for the development of low-cost, sensor-equipped aircraft that can operate in previously unreachable atmospheric layers at altitudes of 55-80 km on Earth and 20-40 km on Mars, opening new avenues for a deeper understanding of Earth's atmosphere as well as exploration of other planetary atmospheres.

**Acknowledgements:** This work was supported in part by NASA's NSGRTO (80NSSC20K1191) and NIAC (80NSSC23K0590) programs.

# 3D photophoretic aircraft made from ultralight porous materials can carry kg-scale payloads in the mesosphere
Supplementary Information
Thomas Celenza, Andy Eskenazi and Igor Bargatin

In this document, we present and expand on the computational and theoretical framework behind our work. The first section is devoted to the ANSYS Fluent simulations, covering the solver set-up and the theory behind the force calculations. The second section of this document focuses on the MATLAB code, specifically the derivation of the equations used in the optimization of the geometrical and channel parameters of the 3D geometries, including the rocket, cone and sphere. Finally, the third section expands on buckling simulations conducted in COMSOL.

## 1. ANSYS Fluent Simulations

The goal of the ANSYS Fluent simulations was to determine an analytical expression to estimate the lift forces produced by various types of 3D structures. The specific solver implemented was the SST k-omega solver with default settings except the low-Re number corrections option. Because we sought geometries that operated across a wide range of velocities and altitudes (and thus air pressures, densities, temperatures and viscosities), the expression for the lift force needed to be valid across a wide range of Reynolds (Re) numbers as well. In particular, this equation needed to reasonably accurately model the transition between the low-Re (Stokes) regime to the high-Re regime. As the main paper argues, an appropriate expression is

$$F = C_1 8\mu D v_{ft} + C_2 \rho K v_{\text{jet}}^2 \,. \tag{S1}$$

Here, $\mu$ corresponds to the fluid viscosity, $\rho$ to the density, $K = \pi r^2$ is the area of a nozzle with radius $r$, $D$ is the geometry's characteristic (usually largest) dimension, while $v_{ft}$ is the flow-through velocity of the fluid through the porous material and $v_{jet}$ is the velocity of the fluid exiting the structure through the nozzle.

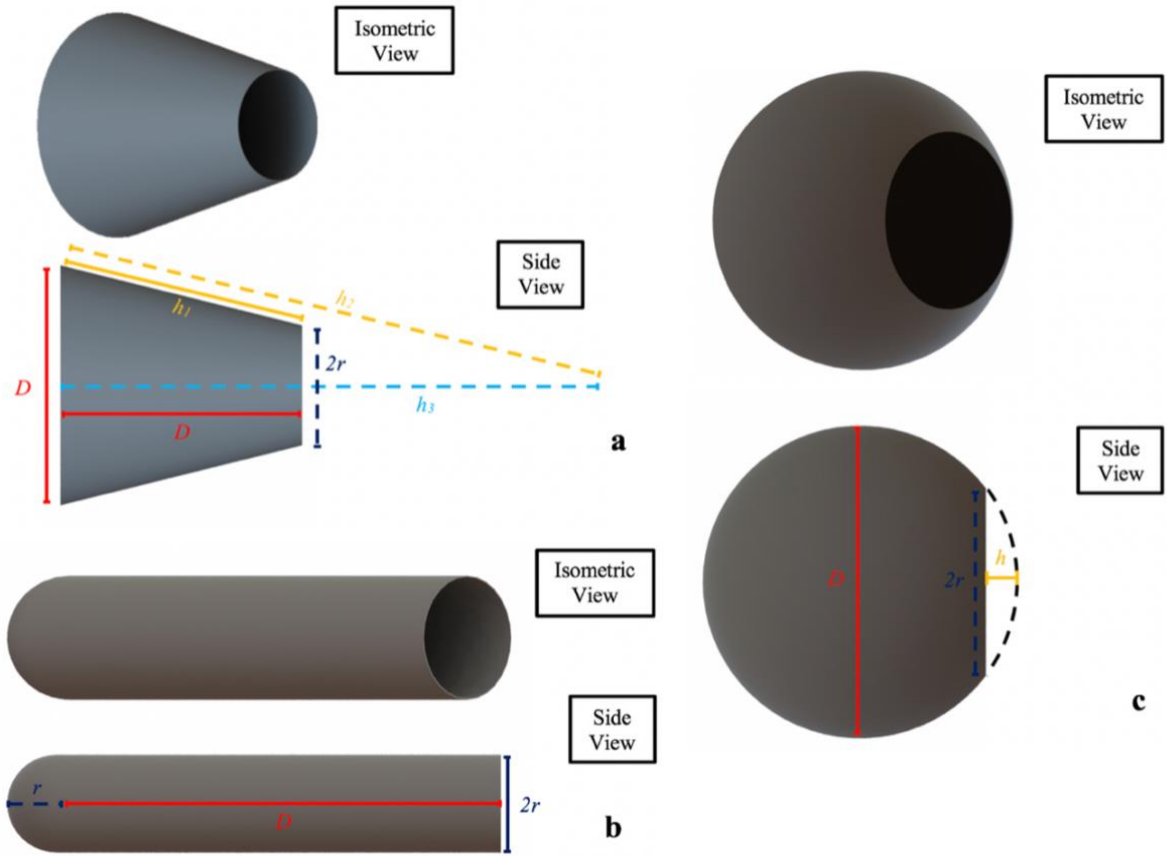

***Figure S1***: *Main geometric parameters for the cone (a), rocket (b) and sphere (c). Notice that here, the variable D serves as an overall indicator of the size of the geometry, while the variable r controls the outlet radii of the nozzle.*

The fitting parameters $C_1$ and $C_2$ depended on the geometry and were determined using ANSYS simulations. In this work, we considered three geometries, a cone, sphere, and rocket, shown in **Fig. S1**. Through the ANSYS simulations, we determined the average $C_1$ and $C_2$ coefficients for each structure and examined how these would evolve with overall size of the structure or the altitude. We performed 9 sets of simulations for each geometry, where we varied three different inlet/outlet area ratios at three different altitudes, resulting in flow-through velocities as small as $10^{-6}$ m/s or as large as 1 m/s.

To make our simulations computationally efficient, we took advantage of the axial symmetry of our three geometries and thus constructed our models in a 2D, axisymmetric environment, which allowed us to only simulate fluid flow on the top half of each structure. We formed these geometries using ANSYS' "Design Modeler" module, and they were essentially composed of three spaces: an outer air box, and inner air box, and the nanocardboard geometry itself (whose interior was "subtracted" from the inner air box). For the purposes of these simulations and the limitations posed by Fluent, we assumed 100% porous walls through which the air would flow at velocity $v_\text{ft}$ (an idealization of the actual porous nanocardboard material our geometries were made of). We believe this assumption to be valid because we were mainly investigating overall geometric effects created by changing the shape. Furthermore, although real nanocardboard, as demonstrated in previous work [R5], had a porosity of approximately 50%, the gas flowing through each of its pores would do so about twice as fast than the flow-through velocity $v_\text{ft}$ in our simulations. This is because we defined the flowthrough velocity to be the volumetric flow rate of the gas through the pores divided by the total area of nanocardboard (rather than just the area of the pores). As a result, changing the porosity would not affect the obtained fitting coefficients $C_1$ and $C_2$ since the reaction forces acting on the flyer are calculated from the force acting on the air box, which is not dependent on the total volume of air flow into the flyer and out of the jet.

The next step was to specify mesh elements, shown in **Fig. S2**. Plot (a) shows the larger, outer air box with coarser mesh elements, while plot (b) is a zoomed-in view into the smaller, inner air box, containing smaller mesh elements. By dividing the air box into these two regions, we optimized the overall number of mesh elements in the simulation by providing a higher resolution just in the area close to the geometry. We created the mesh by selecting edges and dividing them into a discrete number of points; to enforce a uniform grid pattern, we used the quadrilaterals face meshing command. For the sphere, this resulted in 184,180 elements (185,408 nodes); for the cone, 194,322 elements (195,865 nodes); for the rocket, 293,053 elements (294,616 nodes). These were the final numbers of mesh elements obtained as a result of performing a convergence analysis until observing negligible changes in the computed lift forces.

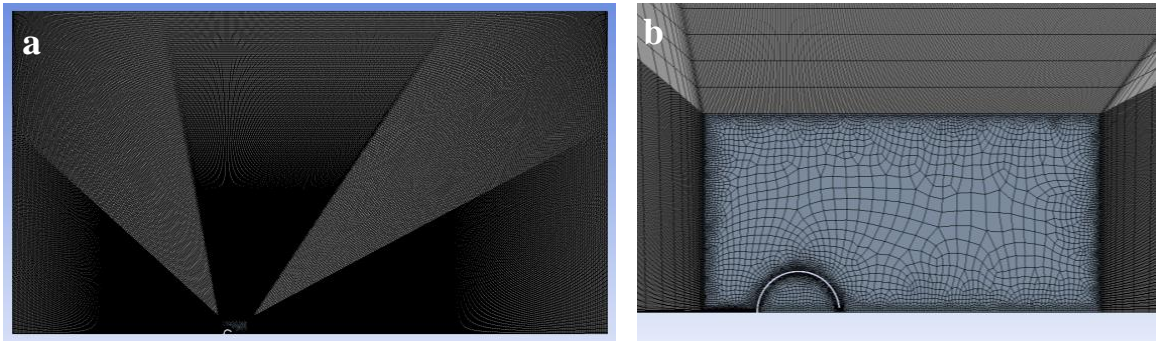

*Figure S2*: Sample meshing of the axisymmetric sphere simulation in ANSYS Fluent. Here, plot (a) provides an overall picture of the air box (which is more than ten times larger than the geometry in question in each dimension), while plot (b) shows a zoomed-in image of the area immediately surrounding the sphere. The size of the outer air box was not arbitrary, but rather resulted from a series of simulations that gradually increased its dimensions until force values converged.

The final step was to establish Fluent's "set-up" module parameters. For the model, we chose the viscous k-omega, with the low-Re (viscous) corrections feature enabled. Next, we fixed the boundary conditions as described in the main paper, and manually modified operating conditions (environment pressure, fluid density and fluid viscosity) matching the chosen altitude. Since our fluid was air, we extracted its properties as tabulated in altitude-dependent standard atmospheric tables, summarized in **Table 1** below for 0 km, 40 km and 70 km (our probed altitudes). Last, we specified the inlet velocity as a variable parameter, since that allowed us to sweep through values ranging from $10^{-6}$ m/s to 1 m/s in 7 logarithmically equally spaced points.

| Summary of Altitude-Dependent Atmospheric Properties | | | |
|---|---|---|---|
| Altitude | 0 km | 40 km | 70 km |
| Atmospheric Pressure (Pa) | 101300 | 275.47 | 4.66 |
| Atmospheric Temperature (K) | 288 | 251 | 220 |
| Air Density (kg/m3) | 1.23 | $3.83*10^{-3}$ | $7.38*10^{-5}$ |
| Air Viscosity (Pa * s) | $1.796*10^{-5}$ | $1.610*10^{-5}$ | $1.447*10^{-5}$ |

*Table 1*: Tabulated altitude-dependent atmospheric conditions for 0 km, 40 km and 70 km. These values were manually inputted for each simulation set into the Fluent solver.

We repeated this process 36 times, to construct 18 simulations for the cone, 9 for the sphere and 9 for the rocket, using operating conditions corresponding to 3 different altitudes (0 km, 40 km and 70 km) and 3 different geometry sizes. In each case, we computed the reaction force in the axisymmetric direction using a line integral along the walls of the outer air box, resulting in the force values shown in **Figs. S3–S6**. This computation made use of the fact that under steady-state operation, the reaction force is equal to the lift force. The $C_1$ and $C_2$ coefficients were then determined by fitting the data in MATLAB to equation (S1), resulting in the values that are shown in the same figures and tabulated in **Tables 2-5**. In general, most curves of **Figs. S3–S6** (in the logarithmic scale) show a transition from the viscous, low-Re regime to the high-Re regime that is manifested through a change in the slopes of the force curves. However, at 70 km in altitude, the lift force stayed in the Stokes (low-Re) regime and the high-Re $C_2$ coefficients remained uncertain at this particular altitude. Thus, when computing the overall average $C_1$ and $C_2$, we did not incorporate the $C_2$ corresponding to the 70 km altitude.

| Fitting Parameters for the Rocket, Dia. = 2 cm | | | | | | |
|---|---|---|---|---|---|---|
| Altitude | Length = 1 cm | | Length = 5 cm | | Length = 10 cm | |
| | $C_1$ | $C_2$ | $C_1$ | $C_2$ | $C_1$ | $C_2$ |
| 0 km | 2.0 (1.6–2.4) | 1.1 (0.9–1.3) | 1.0 (0.8–1.2) | 1.1 (0.9–1.2) | 0.9 (0.7–1.1) | 1.1 (0.9–1.2) |
| 40 km | 2.24 (2.12–2.38) | 0.73 (0.62–0.85) | 1.1 (1.0–1.3) | 0.8 (0.6–1.0) | 1.0 (0.9–1.2) | 0.8 (0.6–1.0) |
| 70 km | 2.361 (2.353–2.368) | | 1.20 (1.20–1.20) | | 1.08 (1.08–1.10) | |
| Average | 2.22 | 0.91 | 1.12 | 0.92 | 1.00 | 0.95 |

*Table 2*: $C_1$ and $C_2$ coefficients computed for the rocket geometry of different lengths (1 cm, 5 cm and 10 cm), alongside the 66% confidence intervals for each fitting parameter (tabulated below each coefficient entry).

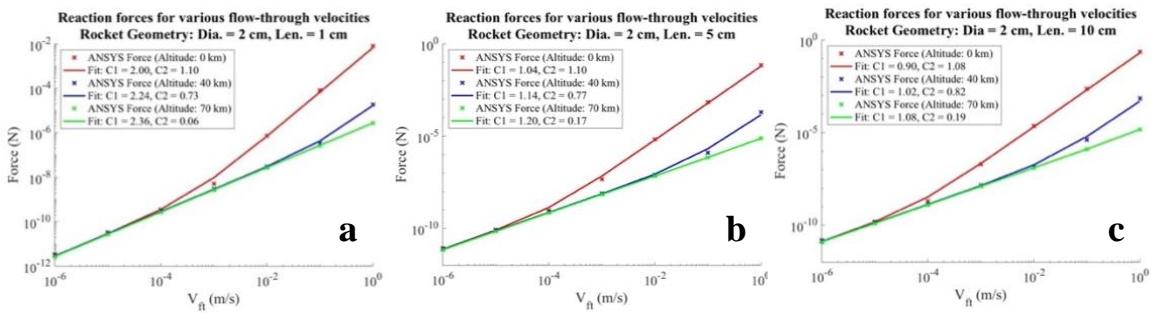

*Figure S3*: Results from the altitude-dependent rocket simulations in ANSYS Fluent; each data point corresponds to a different flow-through velocity, ranging from $10^{-6}$ m/s to 1 m/s, while plots (a), (b) and (c) correspond to different rocket lengths.

| Fitting Parameters for the Sphere, Dia. = 2 cm | | | | | | |
|---|---|---|---|---|---|---|
| Altitude | $r_{out}$ = 0.1 cm | | $r_{out}$ = 0.5 cm | | $r_{out}$ = 1 cm | |
| | $C_1$ | $C_2$ | $C_1$ | $C_2$ | $C_1$ | $C_2$ |
| 0 km | **1.4** (0.7–2.0) | **0.29** (0.21–0.37) | **1.5** (1.3–1.7) | **1.06** (0.95–1.18) | **0.9** (0.8–1.0) | **1.5** (1.4–1.7) |
| 40 km | **1.4** (1.0–1.9) | **0.6** (0.4–0.8) | **1.5** (1.3–1.6) | **0.9** (0.7–1.0) | **0.91** (0.89–0.93) | **0.99** (0.91–1.08) |
| 70 km | **1.65** (1.63–1.67) | | **1.58** (1.52–1.64) | | **0.95** (0.94–0.96) | |
| Average | 1.48 | 0.45 | 1.50 | 0.97 | 0.91 | 1.26 |

*Table 3*: $C_1$ and $C_2$ coefficients computed for the sphere geometry of different outlet radii (0.1 cm, 0.5 cm and 1 cm), alongside the 66% confidence intervals for each fitting parameter (tabulated below each coefficient entry).

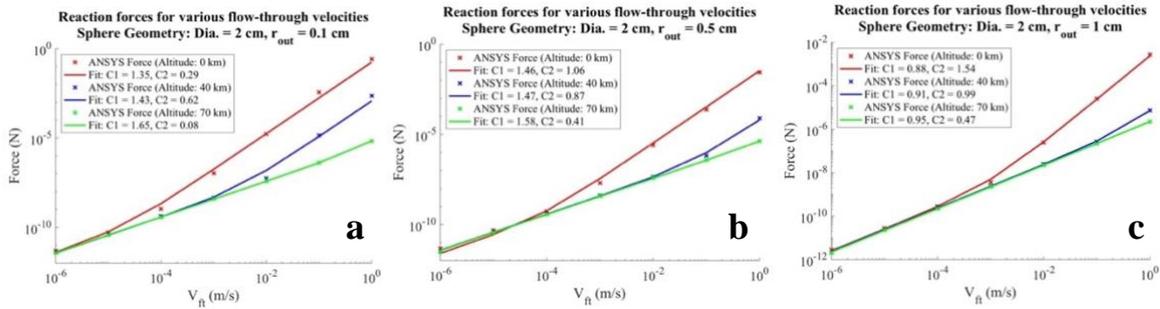

*Figure S4*: Results from the altitude-dependent sphere simulations in ANSYS Fluent; each data point corresponds to a different flow-through velocity, ranging from $10^{-6}$ m/s to 1 m/s, while plots (a), (b) and (c) correspond to different sphere outlet radii.

| Fitting Parameters for the Cone, Dia. = 2 cm | | | | | | |
|---|---|---|---|---|---|---|
| Altitude | Length = 2 cm | | Length = 5 cm | | Length = 10 cm | |
| | $C_1$ | $C_2$ | $C_1$ | $C_2$ | $C_1$ | $C_2$ |
| 0 km | **0.7** (0.5–1.0) | **0.9** (0.7–1.1) | **0.7** (0.5–0.9) | **0.9** (0.8–1.1) | **0.7** (0.4–1.0) | **0.9** (0.7–1.1) |
| 40 km | **1.0** (0.8–1.2) | **0.6** (0.3–0.8) | **0.9** (0.7–1.1) | **0.6** (0.4–0.8) | **0.8** (0.7–1.0) | **0.7** (0.6–0.9) |
| 70 km | **1.07** (0.98–1.16) | | **1.01** (0.94–1.07) | | **0.98** (0.95–1.02) | |
| Average | 0.94 | 0.72 | 0.88 | 0.76 | 0.84 | 0.82 |

*Table 4*: $C_1$ and $C_2$ coefficients computed for the cone geometry (2 cm diameter) of different lengths (2 cm, 5 cm and 10 cm), alongside the 66% confidence intervals for each fitting parameter (tabulated below each coefficient entry).

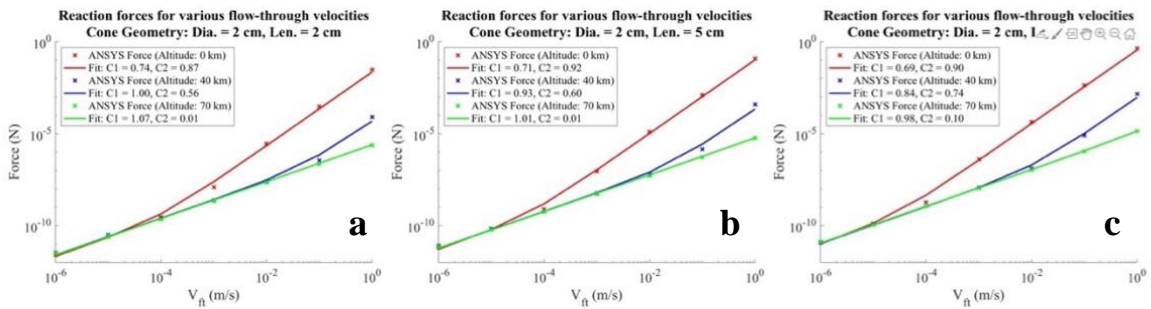

*Figure S5*: Results from the altitude-dependent cone (2 cm diameter) simulations in ANSYS Fluent; each data point corresponds to a different flow-through velocity, ranging from $10^{-6}$ m/s to 1 m/s, while plots (a), (b) and (c) correspond to different cone lengths.

| Fitting Parameters for the Cone, Dia. = 4 cm | | | | | | |
|---|---|---|---|---|---|---|
| Altitude | Length = 2 cm | | Length = 5 cm | | Length = 10 cm | |
| | $C_1$ | $C_2$ | $C_1$ | $C_2$ | $C_1$ | $C_2$ |
| 0 km | **0.9** (0.7–1.1) | **1.0** (0.8–1.2) | **1.0** (0.8–1.2) | **1.0** (0.8–1.1) | **1.0** (0.7–1.3) | **1.0** (0.8–1.1) |
| 40 km | **1.4** (1.1–1.7) | **0.6** (0.4–0.9) | **1.2** (1.0–1.3) | **0.7** (0.6–0.9) | **1.1** (1.0–1.2) | **0.8** (0.7–1.0) |
| 70 km | **1.5** (1.3–1.6) | | **1.24** (1.22–1.25) | | **1.19** (1.18–1.20) | |
| Average | **1.27** | **0.82** | **1.13** | **0.86** | **1.09** | **0.89** |

*Table 5*: $C_1$ and $C_2$ coefficients computed for the cone geometry (4 cm diameter) of different lengths (2 cm, 5 cm and 10 cm), alongside the 66% confidence intervals for each fitting parameter (tabulated below each coefficient entry).

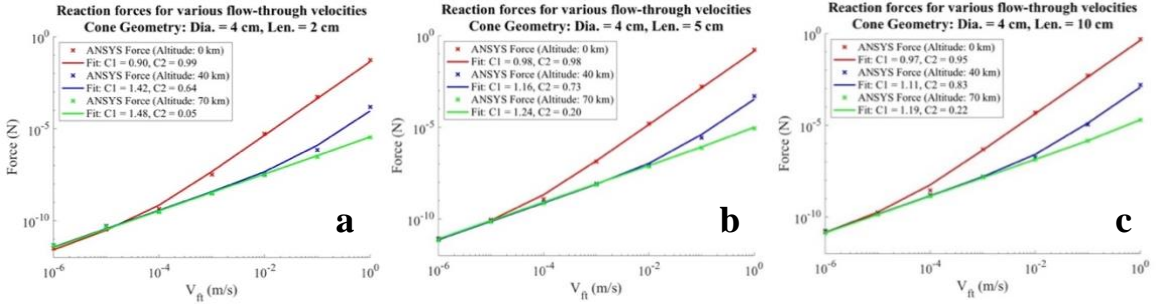

*Figure S6*: Results from the altitude-dependent cone (4 cm diameter) simulations in ANSYS Fluent; each data point corresponds to a different flow-through velocity, ranging from $10^{-6}$ m/s to 1 m/s, while plots (a), (b) and (c) correspond to different cone lengths.

As we increased in altitude, the value of the $C_1$ parameter increased while that of $C_2$ decreased. All in all, **Table 6** below summarizes the average $C_1$ and $C_2$ coefficients obtained for each geometry. In all cases, the coefficients are on the order of 1.

| Average Fitting Parameters for Each Geometry | | | | |
|---|---|---|---|---|
| Geometry | Cone | | Sphere | Rocket |
| | $D$ = 2 cm | $D$ = 4 cm | $D$ = 2 cm | $D$ = 2 cm |
| $C_1$ | 0.9 | 1.2 | 1.3 | 1.4 |
| $C_2$ | 0.8 | 0.9 | 0.9 | 0.9 |

*Table 6*: Fitting parameters for the analytical theory for standard atmospheric conditions on Earth, for each geometry.

The next section of this document takes the force fitting parameters found from the ANSYS Fluent simulations and focuses on MATLAB-based parametric optimization of our three different geometries.

# 2. MATLAB Code and Extension of Theoretical Framework

In this section of the supplementary information, we present the extension to 3D structures of the original nanocardboard fluid mechanic theory developed by [R3]. The equations derived below were implemented in a MATLAB code to perform a series of parametric studies that seek to optimize the geometric and porous parameters of our three study geometries, a cone, a sphere and a rocket. More information about our code can be found in our <u>publicly available repository [R4]</u>.

## 2.1. Derivation of Equations

### 2.1.1 General Overview

For a general 3D porous structure, conservation of mass establishes that

$$K_{total} v_{ft} = K_{out} v_{out} . \tag{S2}$$

Here, $K_{total}$ represents the total surface area of the structure (as if the structure had no pores/channels) and $v_{ft}$ is the flow-through velocity of the fluid across this surface. Similarly, $K_{out}$ corresponds to the area covered by the outlet, while $v_{out}$ is the exit velocity of the fluid out of the structure. Adding Bernoulli's equation, we get the relationship that

$$\frac{P_{in} - P_{out}}{\rho} = \frac{\Delta P}{\rho} = \frac{v_{out}^2 - v_{ft}^2}{2}. \tag{S3}$$

In (S3), $P_{in}$ is the pressure right at the inlet of the structure, $P_{out}$ is the pressure right as the jet of fluid is leaving the structure, located around the space close to $K_{out}$, while $\rho$ is the fluid density. This equation can be rearranged to yield an expression for the pressure difference across both ends of the structure, resulting in

$$\Delta P = \frac{\rho(v_{out}^2 - v_{ft}^2)}{2}. \tag{S4}$$

Assuming that the porosity of the 3D structure originates from using the nanocardboard geometry developed by [R3] as the wall material, then we can model the mass flow rate of the fluid across one of the structure's pores (or more properly said, channels) using the following equation

$$\dot{m} = -\alpha \Delta P + \gamma \Delta T. \tag{S5}$$

In (S5), $\alpha$ and $\gamma$ represent two constants, which come from curve-fitting the data obtained by [R7] and transforming the non-dimensional mass flow rate equation into a dimensional form again, with both pressure and temperature contributions[1]. These variables take the following form:

$$\alpha = \left(\frac{\delta}{6} + 1\right)\left(1 + \frac{0.25}{\sqrt{\delta}}\right)\frac{A^2 B \beta_*}{L}, \tag{S6}$$

and

$$\gamma = \left(\frac{1.1}{1.5 + \delta}\right)\frac{A^2 B P_* \beta_*}{T_* L}. \tag{S7}$$

Here, the variable $P_*$ denotes the average pressure[2] between the two sides of the structure's nanocardboard wall, $T_*$ analogously describes the average temperature between both sides of the wall's surface, while $\beta_*$ is an inverse velocity parameter. Specifically, this last one is given by

$$\beta_* = \sqrt{\frac{m}{2k_B T_*}}, \tag{S8}$$

where $k_B$ is the Boltzmann constant (equal to $1.38 * 10^{-23}$ J/K), and $m$ is the mass of an air molecule[3]. Lastly, the parameter $\delta$ is the gas rarefaction coefficient, which [R7] defines as

$$\delta = \frac{\sqrt{\pi} A}{2\lambda} = \frac{\sqrt{\pi}}{2 Kn}. \tag{S9}$$

In this expression, $\lambda$ is the molecular mean free path, defined as the average distance traveled by a molecule between collisions with other molecules, and $Kn$ is the Knudsen number, which is characterized in terms the of channel width. In essence, higher values of the $\delta$ parameter designates flows in the continuum regime, while smaller values indicate flows taking place in the free molecular regime. As for the molecular mean free path, mathematically it is usually expressed as

$$\lambda = \frac{\mu(T)}{P(T)}\sqrt{\frac{\pi k_B T}{2m}} = \frac{\mu(T)}{P(T)}\sqrt{\frac{\pi R_{air} T}{2}},\tag{S10}$$

where $\mu(T)$ is the fluid's viscosity and $P(T)$ is the operating pressure, both given as a function of $T$, the operating temperature. In addition, from equation (S9), we see the Knudsen number is defined as

$$Kn = \frac{\lambda}{A}.\tag{S11}$$

---

[1] For more information, please see [R2].
[2] The value of this variable may be found from performing CFD simulations but will be simply approximated as the operating pressure.
[3] The molar mass of air is 0.02896 kg/mol, so then the approximated mass of an air molecule would be $0.02896/(6.022*10^{23})$ (Avogadro's number), or $4.8089 * 10^{-26}$ kg.

Additionally, as seen in **Fig. S7** below, the variables $A$ and $B$ characterize the nanocardboard channel's width and length, respectively, yielding a cross-sectional area of $A$ x $B$. In addition, $L$ denotes the channel's height. Note that in [R3], $A$ is assumed to be much smaller than $B$.

After defining these variables and introducing the expression for the mass flow rate, $\dot{m}$, across one of nanocardboard's channels, then an equation can be derived for the average flow-through velocity across the structure's surface, which is simply described by

$$v_{ft} = \frac{\varphi \dot{m}}{\rho AB} = \frac{\varphi(-\alpha \Delta P + \gamma \Delta T)}{\rho AB}.\tag{S12}$$

Here, $\dot{m}/\rho$ is no other than the volumetric flow rate $\dot{V}$, while the term $\varphi$ denotes the geometric fill factor, which is defined in terms of $K_{in}$ (porous area) and $K_{total}$[4], or the channel parameters, and takes the form

$$\varphi = \frac{K_{in}}{K_{total}} = \frac{ABX}{(ABX + SBX)} = \frac{A}{(A+S)}.\tag{S13}$$

The latter two equivalencies in (S13) originates from analyzing a single nanocardboard unit cell as opposed to the full 3D structure. Indeed, as **Fig. S7** shows, the total cross-sectional area of the cell (if no channels were present) is given by

$$K_{cell} = (ABX + SBX) = (A+S)BX,\tag{S14}$$

where the variable $X$ is just the number of channels in a unit cell.

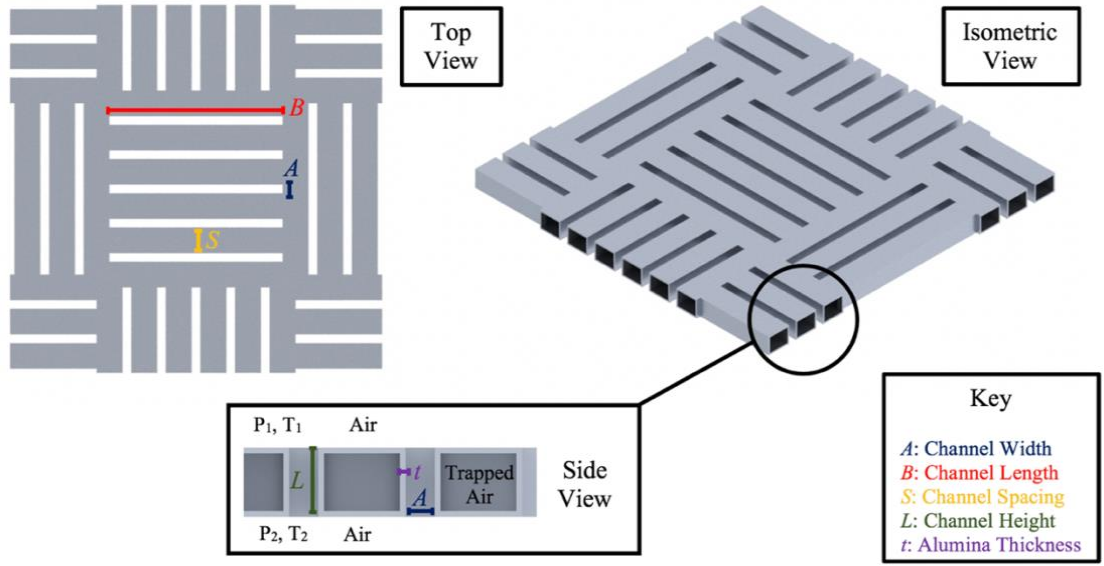

*Figure S7: Main nanocardboard channel parameters.*

---
[4] This area is essentially the total 3D structure wall area if there were no channels present. This is analogous to $Area_{cell}$ in the single nanocardboard unit cell.

However, this number (*X*) is not arbitrarily chosen, and is dictated by *A, B* and *S* in the following way

$$X = \frac{B - S}{S + A}. \tag{S15}$$

This expression considers the channel width *A* and spacing *S* as a unit, and tries to fit as many of those *A* + *S* units into the channel length *B*. Nonetheless, we need to consider an additional *S* for spacing against the perpendicular channels. This can be seen more clearly in **Fig. S8** below, where the yellow bars represent the *A* + *S* units, and as drawn, five of these fit in the length of *B*, after subtracting one *S*.

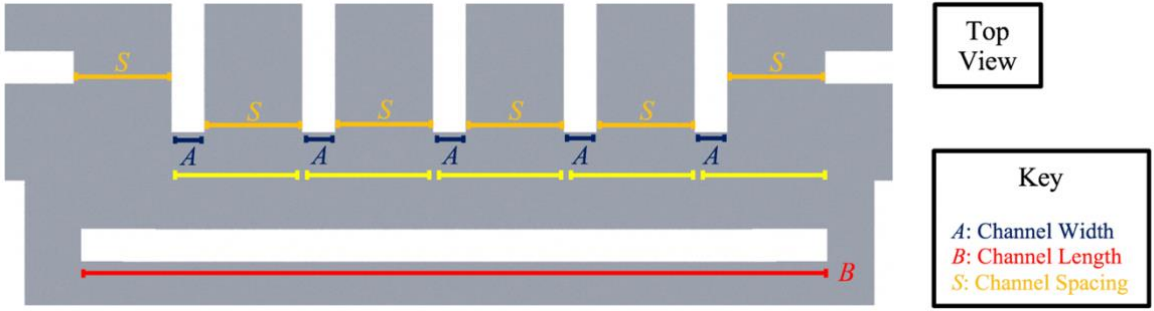

***Figure S8**: Illustration of equation (S15), with the yellow bars showing the A + S units fitted into the channel length B.*

Overall, the flow-through velocity expression provided in (S12) is a step closer towards calculating the lift force that a 3D structure could generate for a given combination of geometric and channel parameters. However, computing lift will not be possible until we solve for $v_{out}$. Therefore, (S12) can be rearranged to instead solve for another unknown, $\Delta P$, and obtain

$$\Delta P = \frac{\gamma \Delta T}{\alpha} - \frac{v_{ft} \rho A B}{\alpha \varphi}. \tag{S16}$$

Since both (S16) and (S4) from above provide two distinct expressions for the pressure difference, it is possible to equate them, giving rise to yet another relationship between $v_{ft}$ and $v_{out}$, giving

$$\frac{\rho(v_{out}^2 - v_{ft}^2)}{2} = \Delta P = \frac{\gamma \Delta T}{\alpha} - \frac{v_{ft} \rho A B}{\alpha \varphi}. \tag{S17}$$

Rearranging this expression further, we get

$$v_{out}^2 = \frac{2}{\rho}\left(\frac{\gamma \Delta T}{\alpha} - \frac{v_{ft} \rho A B}{\alpha \varphi}\right) + v_{ft}^2. \tag{S18}$$

Now, recalling the conservation of mass relationship provided in (S2), it is possible to write $v_{ft}$, the flow-through velocity across the channels, in terms of $v_{out}$

$$v_{ft} = \frac{K_{out}}{K_{total}} v_{out} = \frac{\varphi K_{out}}{K_{in}} v_{out}. \tag{S19}$$

Thus, (S19) can replace the $v_{ft}$ term in (S18), leaving everything in terms of just $v_{out}$

$$v_{out}^2 = \frac{2}{\rho}\left(\frac{\gamma \Delta T}{\alpha} - \frac{K_{out} v_{out} \rho A B}{K_{total} \alpha \varphi}\right) + \left(\frac{K_{out}}{K_{total}}\right)^2 v_{out}^2. \tag{S20}$$

Further manipulating (S20), we get the following quadratic

$$v_{out}^2 \left(1 - \left(\frac{K_{out}}{K_{total}}\right)^2\right) + v_{out}\left(\frac{2 K_{out} A B}{K_{total} \alpha \varphi}\right) - \frac{2 \gamma \Delta T}{\rho \alpha} = 0, \tag{S21}$$

which has precisely $v_{out}$ as its only unknown. This equation will always have two distinct, real solutions, one of which will be positive and the other of which will be negative. These are

$$v_{out}^+ = \frac{-\left(\frac{2K_{out}AB}{K_{total}\alpha\varphi}\right) + \sqrt{\left(\frac{2K_{out}AB}{K_{total}\alpha\varphi}\right)^2 + \frac{8\gamma\Delta T}{\rho\alpha}\left(1-\left(\frac{K_{out}}{K_{total}}\right)^2\right)}}{2\left(1-\left(\frac{K_{out}}{K_{total}}\right)^2\right)} \quad \text{(S22)}$$

as the positive solution, and

$$v_{out}^- = \frac{-\left(\frac{2K_{out}AB}{K_{total}\alpha\varphi}\right) - \sqrt{\left(\frac{2K_{out}AB}{K_{total}\alpha\varphi}\right)^2 + \frac{8\gamma\Delta T}{\rho\alpha}\left(1-\left(\frac{K_{out}}{K_{total}}\right)^2\right)}}{2\left(1-\left(\frac{K_{out}}{K_{total}}\right)^2\right)}. \quad \text{(S23)}$$

In the context of our work, only $v_{out}^+$ is meaningful, since it is the solution that makes physical sense as the outflowing jet of air operates effectively with a positive $v_{out}$. Consequently, throughout the manuscript and the supplementary solution, it is $v_{out}^+$ the solution that is simply referred to as $v_{out}$. A negative solution would involve some mechanism actively pumping the air into the structure through the "nozzle". Since no such pump would exist in practice, we are ignoring the negative solution deeming it non-physical.

One underlying advantage of this derivation was that it removed the need to know the pressure difference, $\Delta P$, while providing us with enough information to solve for $v_{out}$ and $v_{ft}$. In the following sub-section, we deliver more details on the heat conduction modeling across the nanocardboard's thickness, which enabled obtaining an expression for the temperature difference, $\Delta T$, necessary to solve for $v_{out}$ in (S23).

### 2.1.2 Heat Conduction Modeling

#### 2.1.2.1 Full Analytical Derivation for $\Delta T$

In order to compute $\Delta T$, the temperature difference between both sides of the structure's walls, we needed to model the heat conduction across the structure's thickness. We performed a heat energy balance that considered heat transfer across three distinct cross-sectional areas: the channel's column of air, across the alumina thickness of the channel, and across the air trapped within the structure, as shown in **Fig. S9** below. As a result, we can let $Q_t$, the total heat transfer, be

$$Q_t = \frac{\Delta T}{R_{t1}} + \frac{\Delta T}{R_{t2}} + \frac{\Delta T}{R_{t3}}, \quad \text{(S24)}$$

where the $R_{t1}$, $R_{t2}$ and $R_{t3}$ represent the thermal resistances under the three scenarios detailed above. For the first of these areas ($K_1$), the column of air in the channel, we define its thermal resistance as

$$R_{t1} = \frac{L}{k_{air}K_1X} = \frac{L}{k_{air}ABX}, \quad \text{(S25)}$$

where $k_{air}$ is the thermal conductivity of air, $L$ is as usual the channel height, and $ABX$ is the cross-sectional area of the individual channels ($AB$) multiplied by the number of channels ($X$) in a unit cell, as shown in **Fig. S9** above. Notice that $\kappa_{air}$ is both temperature and pressure dependent, as the equation developed by [R10] captures, specifically for the small MEMS scale:

$$\kappa_{air} = \frac{\kappa_0}{\left(1 + \frac{0.00076T}{PL}\right)}. \quad \text{(S26)}$$

In this expression, $\kappa_0$ is the air conductivity at standard atmospheric conditions, normally quoted as $\kappa_0 = 0.024 \frac{W}{m\,K}$. Another comparable and slightly more succinct model for the conductivity of air is from [R8]:

$$\kappa_{air} = \frac{\kappa_0}{\left(1 + \frac{3.116\lambda}{L}\right)} \tag{S27}$$

As the pressure decreases, the mean free path eventually becomes comparable to the channel length, and the effective conductivity starts to decrease below the continuum value. Both equations (S26) and (S27) yielded very similar values for the conductivity of air as a function of the channel thickness $L$, although we used Eq. S27 in the simulations.

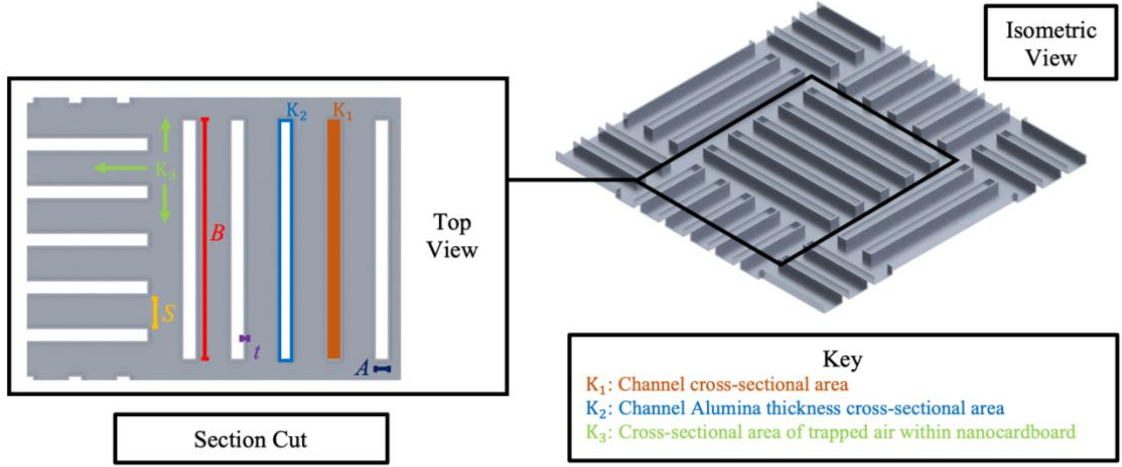

*Figure S9*: Main nanocardboard cross-sectional areas for which thermal resistance is calculated.

Continuing with the heat conduction modeling, the corresponding expression for the thermal resistance across the alumina thickness on the channels (area $K_2$ in **Fig. S9**) is given by

$$R_{t2} = \frac{L}{k_{ald} K_2 X} = \frac{L}{k_{ald}[(A + 2t)(B + 2t) - AB]X}, \tag{S28}$$

where $[(A + 2t)(B + 2t) - AB]X$ is the cross-sectional area occupied by the alumina thickness of the channels, which is denoted as $t$. In (S28), $k_{ald}$ is the thermal conductivity of alumina, which has a constant value of $1.8 \frac{W}{m\,K}$ [R2]. Lastly, the thermal resistance of the air trapped within the structure (area $K_3$) is

$$R_{t3} = \frac{L - 2t}{k_{air} K_3} = \frac{L - 2t}{k_{air}\left[\frac{AB}{\varphi} - (A + 2t)(B + 2t)\right]X}, \tag{S29}$$

where recall from (S13) that $\frac{ABX}{\varphi}$ is the full area of the cell, from which we subtract the combined cross-sectional area of the channels with thickness $t$ of alumina. Now, performing an energy balance, the heat flow through the structure's walls must be equal to that from the absorbed irradiation of the sun, which in this case is given by

$$Q_t = \varepsilon \psi I_{sun} \left(\frac{ABX}{\varphi}\right)(1 - \varphi). \tag{S30}$$

In equation (S30), $\varepsilon$ denotes the absorption coefficient (approximated to 0.9 based-off the measurements from [R3]), $\psi$ the proportion of absorbed optical flux dissipated upward through the nanocardboard (which is assumed to be 0.5 or 50%), and $I_{sun}$ the intensity of the sun at a particular altitude. In particular, this last term can be modeled using the following equation

$$I_{sun} = 1000 + 3.8h, \tag{S31}$$

where the variable *h* refers to the elevation above sea level in kilometers. Notice that this expression returns the sun's intensity in units of Watts per meter square. Furthermore, in equation (S30), $(ABX/\varphi)(1 - \varphi)$ corresponds to the solid area of the nanocardboard, $K_{solid}$, where the sun's irradiation is absorbed. In any case, (S24) through (S31) were combined to write a general expression for $\Delta T$, which is summarized by

$$\Delta T = T_2 - T_1 = \frac{\varepsilon \psi I_{sun}\left(\frac{ABX}{\varphi}\right)(1 - \varphi)}{\frac{1}{R_{t1}} + \frac{1}{R_{t2}} + \frac{1}{R_{t3}}} =$$

$$= \frac{\varepsilon \psi I_{sun}\left(\frac{ABX}{\varphi}\right)(1 - \varphi)}{\frac{k_{air}ABX}{L} + \frac{k_{ald}[(A + 2t)(B + 2t) - AB]X}{L} + \frac{k_{air}\left[\frac{AB}{\varphi} - (A + 2t)(B + 2t)\right]X}{L - 2t}}.$$

(S32)

In (S32), $T_1$ and $T_2$ represent the average temperatures outside and inside the 3D structure, respectively. However, these might not necessarily be known beforehand, reason why calculating $\Delta T$ or $T_*$, the average temperature between both sides of the surface, may not be as trivial. In particular, to compute $T_*$, we make use of the fact that we know what $\Delta T$ is from (S32) and take the following expression

$$T_* = \frac{T_1 + T_2}{2} = \frac{(T_2 - T_1) + 2T_1}{2} = \frac{\Delta T + 2T_1}{2}.$$

(S33)

Here, notice that $T_1$ is simply equal to the temperature corresponding to the particular operating conditions (altitude, pressure, density) of the fluid. Overall, $\Delta T$ allows us to solve for $T_*$ (which is needed to compute $\gamma$ and $\beta_*$ in (S7) and (S9), respectively) and the last part of the puzzle in (S23), the $v_{out}$ expression.

### 2.1.2.2 Simplified Expression for $\Delta T$ in the limit of zero alumina thickness

Beyond the derivation provided in 1.2.1, notice that one could potentially also approximate $\Delta T$ through a more simplified expression given by

$$\Delta T \sim \frac{L I_{sun}(1 - \varphi)}{2\kappa_{air}}.$$

(S34)

The origin of (S34) comes from taking the limit as *t*, the alumina thickness, approaches zero, in equation (S32). Indeed,

$$\lim_{t \to 0} \frac{\varepsilon \psi I_{sun}\left(\frac{ABX}{\varphi}\right)(1 - \varphi)}{\frac{k_{air}ABX}{L} + \frac{k_{ald}[(A + 2t)(B + 2t) - AB]X}{L} + \frac{k_{air}\left[\frac{AB}{\varphi} - (A + 2t)(B + 2t)\right]X}{L - 2t}}$$

$$= \lim_{t \to 0} \frac{L\varepsilon \psi I_{sun}\left(\frac{ABX}{\varphi}\right)(1 - \varphi)}{k_{air}ABX + k_{ald}[(A + 2t)(B + 2t) - AB]X + k_{air}\left[\frac{AB}{\varphi} - (A + 2t)(B + 2t)\right]X}$$

$$= \lim_{t \to 0} \frac{L\varepsilon \psi I_{sun}\left(\frac{ABX}{\varphi}\right)(1 - \varphi)}{k_{air}ABX + k_{ald}[AB - AB]X + k_{air}\left[\frac{AB}{\varphi} - AB\right]X}$$

$$= \lim_{t \to 0} \frac{L\varepsilon \psi I_{sun}\left(\frac{ABX}{\varphi}\right)(1 - \varphi)}{k_{air}ABX + k_{air}\frac{ABX}{\varphi} - k_{air}ABX} = \frac{L\varepsilon \psi I_{sun}\left(\frac{ABX}{\varphi}\right)(1 - \varphi)}{k_{air}\frac{ABX}{\varphi}}$$

(S35)

$$= \frac{L\varepsilon\psi I_{sun}(1-\varphi)}{k_{air}}.$$

Furthermore, letting $\varepsilon = 1$ and $\psi = 0.5$, then (S37) indeed becomes equation (S34) from above. As evidenced by its compressed form, using (S36) to approximate $\Delta T$ simplifies the process of solving for the flow-through velocity, $v_{ft}$. This is especially true if we were to also neglect the pressure term, assuming its contribution is negligible. As a result, the mass flow rate from (S5) can be re-written as

$$\dot{m} \sim \gamma * \Delta T. \tag{S36}$$

This helps reduce the flow-through velocity expression to

$$v_{ft} = \frac{\varphi \dot{m}}{\rho AB} = \frac{\varphi \gamma \Delta T}{\rho AB} = \frac{\varphi \gamma}{\rho AB}\frac{L I_{sun}(1-\varphi)}{2\kappa_{air}}. \tag{S37}$$

Even this expression can be further simplified by reducing the $\gamma$ term from (S7) to

$$\gamma \sim \frac{1.1 A^2 B P_* \beta_*}{\delta T_* L} = \frac{1.1 A^2 B P \beta_*}{TLA\sqrt{\pi}/(2\lambda)} = \frac{2.2\lambda ABP}{\sqrt{\pi}TL}\sqrt{\frac{m}{2k_B T}}. \tag{S38}$$

From the ideal gas law, we have that $P = \rho R_{air} T$, so the pressure term can be replaced in (S38) to obtain

$$\gamma \sim \frac{2.2\lambda AB\rho R_{air} T}{\sqrt{\pi}TL}\sqrt{\frac{m}{2k_B T}} = \frac{2.2\lambda AB\rho R_{air}}{\sqrt{\pi}L}\sqrt{\frac{m}{2k_B T}}. \tag{S39}$$

Combining equations (S37) and (S39), the resultant expression turns out as

$$v_{ft} = \frac{\varphi}{\rho AB}\frac{LI(1-\varphi)}{2\kappa_{air}}\frac{2.2\lambda AB\rho R_{air}}{\sqrt{\pi}L}\sqrt{\frac{m}{2k_B T}} = \frac{1.1\varphi I(1-\varphi)\lambda R_{air}}{\kappa_{air}}\sqrt{\frac{m}{2k_B T\pi}} \tag{S40}$$

Now, recall that the average molecular velocity is equal to

$$v_{avg} = \sqrt{\frac{8 R_{air} T}{\pi}}, \tag{S41}$$

and the relationship between viscosity and velocity, as provided by [R6], is equal to

$$\mu = \frac{\lambda \rho v_{avg}}{2}. \tag{S42}$$

Hence, combining both (S41) and (S42) and solving for $\lambda$, we obtain an expression which can be incorporated in (S40) to yield

$$v_{ft} = \frac{1.1\varphi I(1-\varphi)R_{air}}{\kappa_{air}}\frac{\mu}{P}\sqrt{\frac{\pi k_B T}{2m}}\sqrt{\frac{m}{2k_B T\pi}} = \frac{1.1\varphi I(1-\varphi)R_{air}}{\kappa_{air}}\frac{\mu}{P}\sqrt{\frac{m\pi k_B T}{4k_B T\pi m}}$$

$$= \frac{1.1\varphi I(1-\varphi)R_{air}}{\kappa_{air}}\frac{\mu}{P}\sqrt{\frac{1}{4}} = \frac{1.1\varphi I(1-\varphi)R_{air}}{2\kappa_{air}}\frac{\mu}{P}. \tag{S43}$$

Now, according to [R6], the conductivity of air can be often approximated as $\kappa_{air} = \frac{2\mu C_v'}{M} = 2\mu C_v$, where $M$ is the molar mass of air and $C_v'$ is the specific heat capacity at constant volume, in units of J/k mol. Thus, equation (S43) can further simplify into

$$v_{ft} = \frac{1.1\varphi I(1-\varphi)MR_{air}}{4\mu C_v{'}}\frac{\mu}{P} = \frac{1.1\varphi I(1-\varphi)R_{air}}{4PC_v} = \frac{1.1\varphi(1-\varphi)R_{air}}{4C_v}\frac{I}{P} = C\frac{I}{P}. \quad (S44)$$

where the constant *C* is simply given by

$$C = \frac{1.1\varphi(1-\varphi)R_{air}}{4C_v} = \frac{1.1 \cdot 0.5 \cdot (1-0.5) \cdot 0.287}{4 \cdot 0.718} = 0.0275. \quad (S45)$$

Hence, what these series of derivations shows is that it is possible to approximate and obtain order-of-magnitude estimations of the flow-through velocity by using

$$v_{ft} = 0.0275\frac{I}{P}. \quad (S46)$$

## 2.2. Lift-Force Calculations and Temperature-dependencies

Once we knew how to calculate $v_{ft}$ and $v_{out}$ using the equations derived above (whether it is in the simplified or full analytical form), we used the following equation to calculate the lift forces produced by each geometry, as outlined in the ANSYS simulations section at the beginning of this document:

$$\sum F = C_1 8\mu D v_{ft} + C_2 \rho K_{out} v_{out}^2. \quad (S47)$$

Here, *D* is the characteristic radius of the geometry (usually the inlet radius), while $\mu$ is the viscosity and $\rho$ the fluid density. In addition, $C_1$ and $C_2$ are the geometry dependent coefficients summarized in **Table 6**.

As the derivation of equations above evidences, all of the geometric ($K_{total}$ and $K_{out}$) and channel (*A*, *B*, *L*, *S*, *t*) variables are present in (S23), meaning that it was possible to construct parametric studies exploring the dependency of $v_{ft}$, and consequentially lift, on all of these. Notice, all of these variables were largely independent of each other, making it possible to modify each separately. However, some other parameters within (S23), such as $I_{sun}$, density $\rho$, and air viscosity $\mu$, were actually dependent on temperature, which in turn was also altitude dependent. As a result, in order to accurately calculate the flow-through velocities $v_{ft}$ experienced by a 3D geometry in a range of altitudes, we needed to derive expressions for approximating the air temperature, air pressure, air viscosity and air density as a function of altitude itself.

### 2.2.1 Temperature-dependent Relations

We developed the relations characterizing the dependency between temperature and the fluid variable in question by using standard atmospheric[5] empirical data and fitting equations to it. This approach allowed us to better capture the complex variations of temperature with altitude to a relatively high degree of accuracy; in turn, this process enabled obtaining a more realistic representation of the pressure-altitude dependency, at least compared to typically used approximations such as the barometric formulae. For instance, for the data describing the dependency between air temperature and altitude, we fit both a 6[th], 10[th] and 15[th] order polynomial, as **Fig. S10** to the below shows.

Overall, the 15[th] order polynomial provided the best empirical fit, which was why we decided to use it for the rest of this analysis. However, one interesting aspect of this fit was that we actually fitted the inverse of the temperature, the reason for which will become clearer in the derivation of the altitude-pressure dependency. In any case, equation (S48) below shows this explicit relation, with *h* (the altitude) being in kilometers, and all terms in the column added.

---

[5] The specific standard atmospheric data was taken from the following three websites:
https://www.engineeringtoolbox.com/standard-atmosphere-d_604.html | https://www.pdas.com/atmosTable1SI.html | https://www.pdas.com/bigtables.html

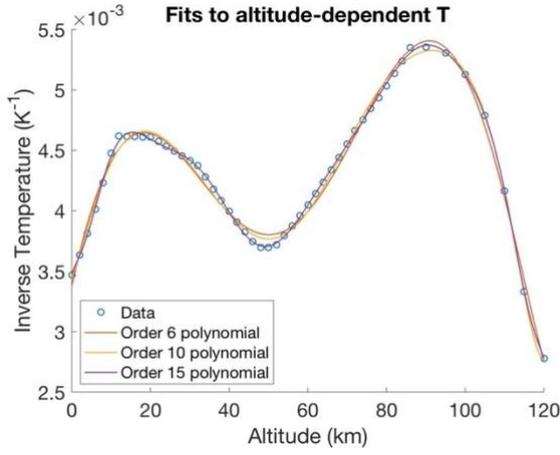

$$T^{-1}(h) = \begin{matrix} -4.592 \cdot 10^{-29} \\ 4.023 \cdot 10^{-27} \\ 1.491 \cdot 10^{-23} \\ -7.942 \cdot 10^{-21} \\ 2.021 \cdot 10^{-18} \\ -3.152 \cdot 10^{-16} \\ 3.271 \cdot 10^{-14} \\ -2.332 \cdot 10^{-12} \\ 1.150 \cdot 10^{-10} \\ -3.862 \cdot 10^{-09} \\ 8.525 \cdot 10^{-08} \\ -1.150 \cdot 10^{-06} \\ 8.154 \cdot 10^{-06} \\ -2.283 \cdot 10^{-05} \\ 9.912 \cdot 10^{-05} \\ 3.473 \cdot 10^{-03} \end{matrix} \cdot \begin{matrix} h^{15} \\ h^{14} \\ h^{13} \\ h^{12} \\ h^{11} \\ h^{10} \\ h^{9} \\ h^{8} \\ h^{7} \\ h^{6} \\ h^{5} \\ h^{4} \\ h^{3} \\ h^{2} \\ h^{1} \\ 1 \end{matrix}. \quad (S48)$$

***Figure S10***: *Modeled temperature dependency on altitude.*

Having derived the empirical relation between temperature (its inverse) and altitude, it was possible to determine a similar expression for pressure. In essence, the differential equation describing the pressure-altitude relationship is given by

$$dP(h) = -g\rho(h) \cdot dh, \quad (S49)$$

where $g$ is the gravitational constant on earth, and $\rho(h)$ the density of air at a particular altitude $h$. Using the ideal gas law, $\rho(h)$ can be substituted to yield the following expression for the above differential in equation (S49)

$$dP(h) = -g \frac{P(h)}{R_{air} T(h)} \cdot dh, \quad (S50)$$

where now $R_{air}$ is the ideal gas constant of air and is equal to $287 \; J/kg * m^3$. Easily enough, one can utilize the technique of separation of variables to obtain that

$$\frac{dP(h)}{P(h)} = \frac{-g}{R_{air} T(h)} \cdot dh, \quad (S51)$$

which leaves all of the pressure terms on one side, and the rest on the other. As a result, it is possible to see with more clarity why the above polynomial fit was done for the inverse of temperature. Indeed, equation (S51) can be equivalently written as

$$\frac{dP(h)}{P(h)} = \frac{-g T^{-1}(h)}{R_{air}} \cdot dh. \quad (S52)$$

This expression can be easily integrated to obtain the following logarithm:

$$\ln(P) = \frac{-g}{R_{air}} \int T^{-1}(h) \cdot dh. \quad (S53)$$

Letting $\zeta(h) = \int T^{-1}(h) * dh$ be a placeholder for the integral of the inverse temperature polynomial and $C$ be simply a constant of integration, we obtain that

$$\ln(P(h)) = \frac{-g}{R_{air}} \zeta(h) + C. \quad (S54)$$

Now, in order to remove the logarithm from the pressure, we can raise both sides of the expression to the Euler's number power, and get

$$P(h) = e^{\frac{-g}{R_{air}} \zeta(h) + C}. \quad (S55)$$

After applying exponent rules, (S55) decomposes into the product given by

$$P(h) = e^C e^{\frac{-g}{R_{air}}\zeta(h)}, \tag{S56}$$

and can be further simplified, upon application of boundary conditions, into

$$P(h) = 101300 \cdot e^{\frac{-g}{R_{air}}*\zeta(h)}, \tag{S57}$$

which takes the following full form:

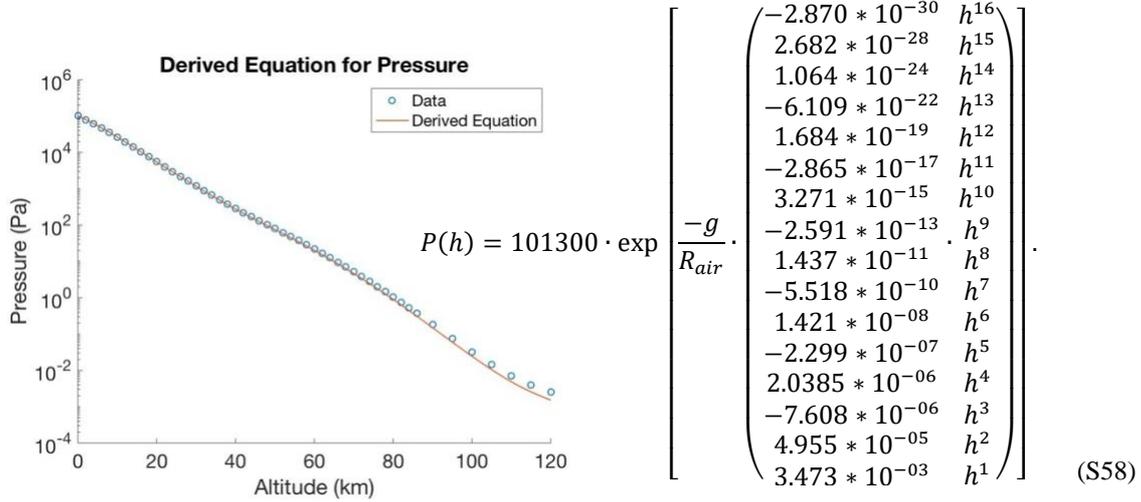

$$P(h) = 101300 \cdot \exp\left[\frac{-g}{R_{air}} \cdot \begin{pmatrix} -2.870*10^{-30} & h^{16} \\ 2.682*10^{-28} & h^{15} \\ 1.064*10^{-24} & h^{14} \\ -6.109*10^{-22} & h^{13} \\ 1.684*10^{-19} & h^{12} \\ -2.865*10^{-17} & h^{11} \\ 3.271*10^{-15} & h^{10} \\ -2.591*10^{-13} & h^{9} \\ 1.437*10^{-11} & h^{8} \\ -5.518*10^{-10} & h^{7} \\ 1.421*10^{-08} & h^{6} \\ -2.299*10^{-07} & h^{5} \\ 2.0385*10^{-06} & h^{4} \\ -7.608*10^{-06} & h^{3} \\ 4.955*10^{-05} & h^{2} \\ 3.473*10^{-03} & h^{1} \end{pmatrix}\right]. \tag{S58}$$

*Figure S11: Modeled pressure dependency on altitude.*

As **Fig. S11** above shows, the agreement of this equation with the empirical data is very reasonable, especially below 80 km altitude. Above 80 km, the atmosphere is no longer well mixed, has increasing concentrations of atomic oxygen, and the simple ideal gas law we used above no longer applies. For this reason, the results that will be presented below correspond to altitudes below 80 km.

The next step was modelling the air density dependency on altitude. With expressions for *T(h)* and *P(h)* above, we could use the ideal gas law to write

$$\rho(h) = \frac{P(h)}{R_{air} T(h)}. \tag{S59}$$

Finally, the last dependency that remained to be defined was the air viscosity and altitude relation. To that end, we could use Sutherland's law, which relates viscosity and temperature through the following equation:

$$\mu(h) = \mu_{ref} \left(\frac{T(h)}{T_{ref}}\right)^{1.5} \left(\frac{T_{ref} + S}{T(h) + S}\right), \tag{S60}$$

where $\mu_{ref}$ is the reference dynamic viscosity and $T_{ref}$ the reference temperature. In this work, for air, at $T_{ref} = 20\ C$, we have that $\mu_{ref} = 0.000018205\ Pa \cdot s$. Finally, $S$ is a constant, known as Sutherland's temperature, which is given by 110.4 K.

### 2.2.2 Payload Calculations

Once all of the required equations and relationships were derived, it was possible to calculate $v_{ft}$ and $v_{out}$ for a specific set of geometric and channel parameters defining unique 3D structures. By calculating these velocities, we determined the total force produced by each geometry, as outlined by equation (S47), from which it was possible to perform some payload estimates. However, in order to obtain the payload estimates, it was paramount to first determine the surface areas of each one of the 3D geometries in question, the reason being that density of these structure was defined in areal terms as opposed to volumetric terms. As was mentioned in the main paper, this work considered a truncated cone, truncated sphere, and a rocket, and their defining equations are shown in **Table 7** below.

| Main Geometrical Area Definitions | | | |
|---|---|---|---|
| Area | Truncated Cone | Truncated Sphere | Rocket |
| $K_{total}$ | $\pi\left(\frac{D}{2}\right)^2 + \pi\left(\frac{D}{2}\right)h_2 - \pi r(h_2 - h_1)$ | $\pi(D^2 - 2rh)$ | $2\pi r(r + D)$ |
| $K_{out}$ | $\pi r^2$ | | |
| $K_{in}$ | $\varphi K_{total}$ | | |
| $K_{solid}$ | $(1 - \varphi)K_{total}$ | | |
| Special Variables | $h_1 = \sqrt{\left(\frac{D}{2} - r\right)^2 + D^2}$ <br> $h_2 = \sqrt{\left(\frac{D}{2}\right)^2 + h^3}$ <br> $h_3 = \frac{D^2}{(D - 2r)}$ | $h = \left(\frac{D}{2}\right) - \sqrt{\left(\frac{D}{2}\right)^2 - r^2}$ | N/A |

***Table 7***: *Area definitions used across this work for the cone, sphere and rocket. Notice that here, the variable $h_3$ follows from using similar triangles analysis, and letting $h_3/(D/2) = D/(D/2 – r)$. For all three geometries, the variable D represents the overall scale of the structure while r their outlet radius. Notice that $K_{in}$ is the porous area, while $K_{solid}$ is the solid area in which the sun's irradiance is absorbed, and it follows that $K_{total} = K_{solid} + K_{in}$.*

As a result, having defined these surface areas (using the parameters established in **Fig. S1**), we calculated the mass of our three 3D structures. In particular, since the cross-sectional area of a channel is simply $AB$, then one can define the number of channels as the following integer floor:

$$n_{channels} = \left\lfloor \frac{K_{in}}{AB} \right\rfloor . \tag{S61}$$

The number of channels, $n_{channels}$, is an important parameter, given that now it is possible to calculate the volume of the structure that is occupied by the deposited alumina around each channel, which has thickness $t$ and relatively high density $\rho_{ald}$ of 3950 kg/m³ [R9]. Indeed, similarly to equation (S28) above, we can define this volume as

$$V_{ald,channels} = n_{channels}(L - 2t)[(A + 2t)(B + 2t) - AB] . \tag{S62}$$

Experimentally, it has already been found that the areal density of nanocardboard, $\sigma_{ncb}$, is about 1 g/m² [R5], but this corresponds to a value of $L$ (nanocardboard thickness) equal to 50 μm. However, in our parametric studies, as we sweep through various values of $L$, especially those that are larger than 50 μm, this areal density alone is not enough to estimate the weight of the structure. As a result, calculating the volume of alumina around each of the channels is paramount, since the structure naturally becomes heavier with increasing thickness. Hence, the overall mass of any one of these geometries will be given by

$$m_{geometry} = \sigma_{geom}(K_{solid} - K_{in}) + \rho_{ald}V_{ald,channels} , \tag{S63}$$

where this expression accounts both for the areal density ($\sigma_{geom}$) and the increases in the amount of the deposited alumina as a result of changes in the wall thickness $L$. Thus, the net lift produced by the geometry is simply given by subtracting the structure's weight from the force expression in (S47), or

$$Lift_{net} = F - gm_{geometry} . \tag{S64}$$

While we know from simulations what $\sigma_{geom}$ is, notice that it is also possible to use our equations and a series of approximations to obtain a theoretical upper bound for this value. In essence, we can start by letting the force be equal to the expression below

$$F = \dot{m}v_{out} = (K_{in}\rho_{air}v_{ft})v_{out} = (K_{in}\frac{P}{R_{air}T}v_{ft})v_{out} \,, \tag{S65}$$

which incorporates mass flow rate and the ideal gas law. Now, recall that equation (S4) provides an expression relating $v_{ft}$ and $v_{out}$, while (S46) provides a simplified approximation for $v_{ft}$. As a result, taking a conservative approach that lets $v_{out} = 0.2 v_{avg}$, a fifth of the average molecular velocity of a gas, shown in (S41) above, and incorporating (S2) and (S46), it is possible to re-write (S68) to obtain

$$F = K_{in}\frac{P}{R_{air}T}v_{ft}0.2\sqrt{\frac{8R_{air}T}{\pi}} =$$
$$= 0.0055 K_{in}\frac{P}{R_{air}T}\frac{I}{P}\sqrt{\frac{8R_{air}T}{\pi}} = 0.0055 K_{in}\frac{I}{R_{air}T}\sqrt{\frac{8R_{air}T}{\pi}}. \tag{S66}$$

Upon further simplification, equation (S69) reduces to

$$F = 0.0055 K_{in}\frac{I}{R_{air}T}\sqrt{\frac{8R_{air}T}{\pi}} = 0.0055\sqrt{\frac{8}{\pi}}K_{in}I\sqrt{\frac{1}{R_{air}T}}. \tag{S67}$$

Thus, the maximum areal density that can be entertained by these 3D structures can be approximated by

$$\sigma_{geom} = \frac{F}{K_{in}g} = 0.0055\sqrt{\frac{8}{\pi}\frac{I}{g}}\sqrt{\frac{1}{R_{air}T}} = KI\sqrt{\frac{1}{R_{air}T}} = 0.016\frac{I}{v_{avg}g} \,, \tag{S68}$$

where $K = \frac{0.0055}{g}\sqrt{\frac{8}{\pi}} = 0.0009$ and $v_{avg} = \sqrt{8R_{air}T/\pi} \approx 400$ m/s is the average velocity of air molecules. Upon inserting the parameters, we find that $\sigma_{geom}$ can have an average value of 0.004 kg/m$^2$, four times of what the areal density of nanocardboard typically is in experiments. The main paper provides additional areal density calculations based off from the parametric studies (detailed below) as well as cloud plots denoting the maximum areal density for each of the study geometries. They are generally of the same order of magnitude as the estimate (S68).

### 2.2.3 Parametric Studies

In this section, we provide four tables that accompany the presentation of the results shown in the main paper. In essence, **Table 8** both summarizes the chosen optimization ranges and discretization for the variables that were varied (*A*, *L* and *r*) and specifies the values that the remaining variables (*B*, *N*, *X*, *S* and *t*) took. Similarly, **Table 9** through **Table 11** present the results for the performed parametric optimization on the three geometries, detailing the specific combination of *A*, *L* and *r* that first, yielded the maximum payload capabilities and second, achieved flight at the lower altitude. In addition, **Table 9** through **Table 11** also provide the areal density of each structure for when the maximum payload was achieved. Notice that this process was repeated for multiple values of *D*, as to explore the dependency of the overall optimization results with the scale of the geometries.

| Parametric Optimization Variables | | | | | |
|---|---|---|---|---|---|
| **Variable** | **Range** | **Truncated Cone** | **Truncated Sphere** | **Rocket** | **Discretization** |
| A | Min. | 10 nm | | | 80 equally spaced points (log scale) |
| | Max. | 5 mm | | | |
| L | Min. | 1 μm | | | 80 equally spaced points (log scale) |
| | Max. | 1 cm | | | |
| r | Min. | $r_{min} = D/20$ (see **Table 9** through **Table 11**) | | | 80 equally spaced points (log scale) |
| | Max. | $r_{max} = D/2.01$ (see **Table 9** through **Table 11**) | | | |
| Altitude | Min. | 0 km | | | 17 equally spaced points (5 km intervals) |
| | Max. | 80 km | | | |
| | | | | | |
| B | | 10A | | | |

| | | |
|---|---|---|
| N | | 1 sun |
| X | | $\dfrac{B - S}{S + A}$ |
| S | | A |
| t | | 50 nm |

*Table 8*: Main values used across the various variables during the parametric optimization. As can be seen, the search range for the optimal A, L and r was discretized in all three cases in 100 points, following a log scale. Changing the granularity of the discretization or the bounds of the search range did not significantly modify the results seen in **Table 9** through **Table 11** below.

| Parametric Optimization Results – Various Sphere Sizes ||||||||
|---|---|---|---|---|---|---|---|
| **Variable** | **Case** | *D* = 2 cm | *D* = 0.1 m | *D* = 0.5 m | *D* = 1 m | *D* = 2 m | *D* = 5 m |
| | | $r_{min}$ = *D*/20, $r_{max}$ = *D*/2.01, with a discretization of 80 points (log scale) ||||||
| A | Max. Payload | 0.90 mm | 0.90 mm | 0.90 mm | 0.90 mm | 0.90 mm | 0.90 mm |
| | Min. Altitude | 0.13 mm | 0.13 mm | 0.20 mm | 0.20 mm | 0.20 mm | 0.20 mm |
| L | Max. Payload | 0.91 mm | 0.91 mm | 0.91 mm | 0.91 mm | 0.91 mm | 0.91 mm |
| | Min. Altitude | 0.14 mm | 0.14 mm | 0.21 mm | 0.21 mm | 0.21 mm | 0.21 mm |
| r | Max. Payload | 9.95 mm | 4.07 cm | 19.05 cm | 36.85 cm | 73.70 cm | 1.84 m |
| | Min. Altitude | 4.05 mm | 1.89 cm | 10.82 cm | 21.63 cm | 43.27 cm | 1.08 m |
| | | | | | | | |
| Max. Payload | Payload (mg) | **8.34** | **79.11** | **1 445** | **5 526** | **21 612** | **133 242** |
| | Altitude (km) | 80 | 80 | 80 | 80 | 80 | 80 |
| | A. Density (g/m²) | 25.48 | 7.81 | 5.91 | 5.64 | 5.54 | 5.49 |
| | Sphere Area (m²) | 0.0007 | 0.025 | 0.64 | 2.63 | 10.52 | 65.82 |
| | $K_{total}/K_{out}$ ratio | 2.22 | 4.77 | 5.68 | 6.17 | 6.17 | 6.17 |
| | | | | | | | |
| Min. Altitude | Payload (mg) | **0.24** | **0.58** | **223.94** | **872.33** | **3 442** | **21 339** |
| | Altitude (km) | 55 | 55 | 60 | 60 | 60 | 60 |
| | $K_{total}/K_{out}$ ratio | 23.34 | 26.96 | 20.30 | 20.32 | 20.31 | 20.38 |

*Table 9*: Combinations of A, L and r that returned the spheres capable of carrying the greatest payload and achieving flight at the lowest altitude, for various values of D, as specified in **Figure S1**.

| Parametric Optimization Results – Various Cone Sizes ||||||||
|---|---|---|---|---|---|---|---|
| **Variable** | **Case** | *D* = 2 cm | *D* = 0.1 m | *D* = 0.5 m | *D* = 1 m | *D* = 2 m | *D* = 5 m |
| | | $r_{min}$ = *D*/20, $r_{max}$ = *D*/2.01, with a discretization of 80 points (log scale) ||||||
| A | Max. Payload | 0.90 mm | 0.90 mm | 0.90 mm | 0.90 mm | 0.90 mm | 0.90 mm |
| | Min. Altitude | 0.13 mm | 0.35 mm | 0.35 mm | 0.35 mm | 0.35 mm | 0.35 mm |
| L | Max. Payload | 0.91 mm | 0.91 mm | 0.91 mm | 0.91 mm | 0.91 mm | 0.91 mm |
| | Min. Altitude | 0.14 mm | 0.36 mm | 0.36 mm | 0.36 mm | 0.36 mm | 0.36 mm |
| r | Max. Payload | 9.95 mm | 4.97 cm | 24.86 cm | 49.73 cm | 99.45 cm | 2.49 m |
| | Min. Altitude | 4.05 mm | 2.39 cm | 11.56 cm | 23.12 cm | 46.25 cm | 1.16 m |
| | | | | | | | |
| Max. Payload | Payload (mg) | **7.96** | **101.26** | **2 043** | **7 929** | **31 228** | **193 348** |
| | Altitude (km) | 80 | 80 | 80 | 80 | 80 | 80 |
| | A. Density (g/m²) | 11.59 | 6.61 | 5.61 | 5.48 | 5.42 | 5.38 |
| | Cone Area (m²) | 0.0016 | 0.039 | 0.98 | 3.92 | 15.67 | 97.97 |
| | $K_{total}/K_{out}$ ratio | 5.04 | 5.05 | 5.05 | 5.04 | 5.04 | 5.03 |
| | | | | | | | |
| Min. Altitude | Payload (mg) | **0.18** | **10.12** | **208.65** | **812.97** | **3 209** | **19 892** |
| | Altitude (km) | 55 | 60 | 60 | 60 | 60 | 60 |
| | $K_{total}/K_{out}$ ratio | 23.97 | 17.75 | 18.84 | 18.84 | 18.83 | 18.72 |

*Table 10*: Combinations of A, L and r that returned the cones capable of carrying the greatest payload and achieving flight at the lowest altitude, for various values of D, as specified in **Figure S1**.

| Parametric Optimization Results – Various Rocket Sizes | | | | | | | |
|---|---|---|---|---|---|---|---|
| **Variable** | **Case** | $D = 2$ cm | $D = 10$ cm | $D = 0.5$ m | $D = 1$ m | $D = 2$ m | $D = 5$ m |
| | | $r_{min} = D/20$, $r_{max} = D/2.01$, with a discretization of 80 points (log scale) | | | | | |
| $A$ | Max. Payload | 0.90 mm | 0.90 mm | 0.90 mm | 0.90 mm | 0.90 mm | 0.90 mm |
| | Min. Altitude | 0.092 mm | 0.13 mm | 0.13 mm | 0.13 mm | 0.13 mm | 0.13 mm |
| $L$ | Max. Payload | 0.91 mm | 0.91 mm | 0.91 mm | 0.91 mm | 0.91 mm | 0.91 mm |
| | Min. Altitude | 0.094 mm | 0.14 mm | 0.14 mm | 0.14 mm | 0.14 mm | 0.14 mm |
| $r$ | Max. Payload | 9.95 mm | 4.97 cm | 24.86 cm | 49.73 cm | 99.45 cm | 2.49 m |
| | Min. Altitude | 1.00 mm | 0.94 cm | 4.12 cm | 8.24 cm | 15.94 cm | 0.40 m |
| Max. Payload | Payload (mg) | **9.51** | **127.59** | **2 639** | **10 281** | **40 573** | **251 516** |
| | Altitude (km) | 80 | 80 | 80 | 80 | 80 | 80 |
| | A. Density (g/m²) | 11.60 | 6.89 | 5.95 | 5.83 | 5.77 | 5.74 |
| | Rocket Area (m²) | 0.0019 | 0.047 | 1.17 | 4.68 | 18.71 | 117.12 |
| | $K_{total}/K_{out}$ ratio | 6.02 | 6.02 | 6.02 | 6.02 | 6.02 | 6.02 |
| Min. Altitude | Payload (mg) | **0.03** | **1.54** | **8.29** | **18.57** | **45.37** | **175.67** |
| | Altitude (km) | 45 | 55 | 55 | 55 | 55 | 55 |
| | $K_{total}/K_{out}$ ratio | 42 | 23.28 | 26.27 | 26.27 | 27.09 | 27 |

*Table 11: Combinations of A, L and r that returned the rockets capable of carrying the greatest payload and achieving flight at the lowest altitude, for various values of D, as specified in Figure S1.*

The results from these tables are discussed in greater detail in the main paper. However, there are four important points to highlight. First, changing $D$ (the scaling of the overall geometries) did not affect significantly the optimal channel parameters $A$ and $L$ that yielded the maximum payload capabilities and achieved flight at the lowest altitude. Secondly, the obtained maximum areal densities were similar across the three geometries (as seen in **Figure 5 (a)** of the main text) and had average values of 9.31 g/m², 6.68 g/m² and 6.96 g/m², for the sphere, cone, and rocket, respectively. Notice that these are above the theoretical order-of-magnitude estimation for the upper limit of 4 g/m² in (S71). Thirdly, the optimized $K_{total}/K_{out}$ ratios (the "Areas Ratio" in the subsequent tables) for the three geometries were relatively invariant across the various values of $D$ and the two missions (max. payload and minimum altitude). For instance, for the maximum payload optimization, $K_{total}/K_{out}$ averaged 5.20, 5.04, and 6.02 for the sphere, cone and rocket, respectively, while for the minimum altitude case, this ratio averaged 21.94, 19.49 and 28.65, respectively. Lastly, for a given surface area, the amount of payload that each geometry could carry was comparable (as illustrated in **Figure 5 (b)** of the main text). As a result, 1 m² of a porous and geometrically optimized cone has a similar maximum payload capability than 1 m² of an optimized rocket and sphere.

Finally, **Fig. S12** through **Fig. S17** present cloud plots that permit visualizing the results from the parametric studies, in particular how different combinations of $A$, $L$ and $r$ enabled geometries with various altitude (a), payload (b) and areal density (c) capabilities. These plots correspond to the $D = 10$ cm and $D = 10$ m cone, sphere and rocket, and are accompanied with illustrations of the optimized geometries that achieved flight at minimum altitude (d) and carried the most payload (e). These figures were generated by discretizing the search ranges of $A$, $L$ and $r$ in 500 equally spaced, and the results from the optimized geometries are shown in Table 12 through Table 14). Despite the increase in discretization points (from 80 to 500) in each dimension, the results were comparable.

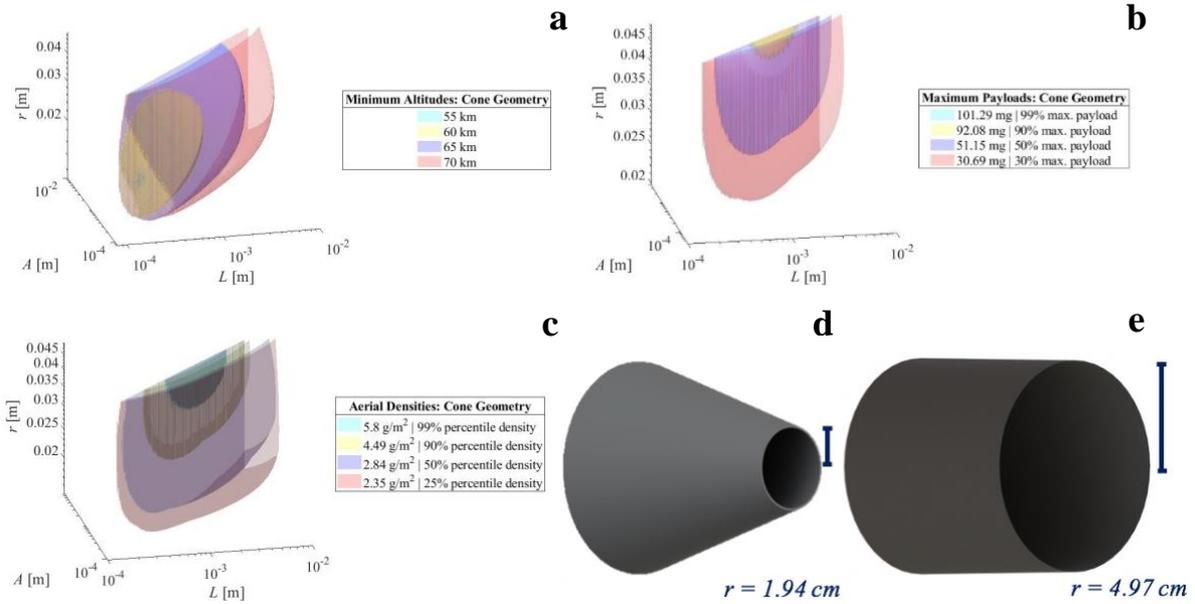

*Figure S12*: Minimum Altitude (a), Maximum Payload (b) and Areal Density (c) plots for the **D = 10 cm Cone Geometry**. Here, the geometry that was able to levitate payload at minimum altitude (0.52 mg at 55 km) is shown in (d), while that which was able to levitate the maximum payload (102.31 mg at 80 km) is shown in (e).

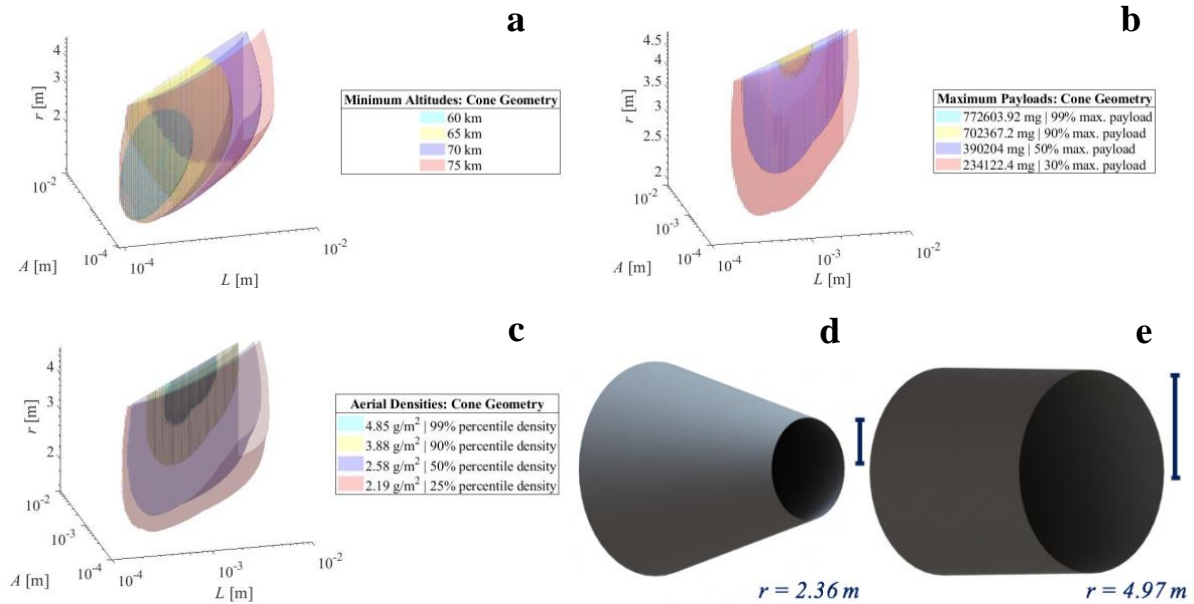

*Figure S13*: Minimum Altitude (a), Maximum Payload (b) and Areal Density (c) plots for the **D = 10 m Cone Geometry**. Here, the geometry that was able to levitate payload at minimum altitude (95 288 mg at 60 km) is shown in (d), while that which was able to levitate the maximum payload (780 408 mg at 80 km) is shown in (e).

| Comparison of D = 10 cm and D = 10 m Cone Geometries | | | | | | | |
|---|---|---|---|---|---|---|---|
| Case | | A | L | r | Surface Area (m²) | Areas Ratio | Payload (mg) | Altitude (km) |
| | | Discretization of 500 points | | | | | | |
| D = 10 cm | Min. Altitude | 0.15 mm | 0.16 mm | 1.94 cm | 0.03 | 25.92 | 0.52 | 55 |
| | Max. Payload | 1.24 mm | 1.25 mm | 4.97 cm | 0.04 | 5.05 | 102.31 | 80 |
| D = 10 m | Min. Altitude | 0.21 mm | 0.22 mm | 2.36 m | 317.52 | 18.16 | 95 288 | 60 |
| | Max. Payload | 1.24 mm | 1.25 mm | 4.97 m | 391.56 | 5.05 | 780 408 | 80 |

*Table 12*: Combinations of A, L and r that returned the optimal cone geometries described in **Figure S12** and **Figure S13** above.

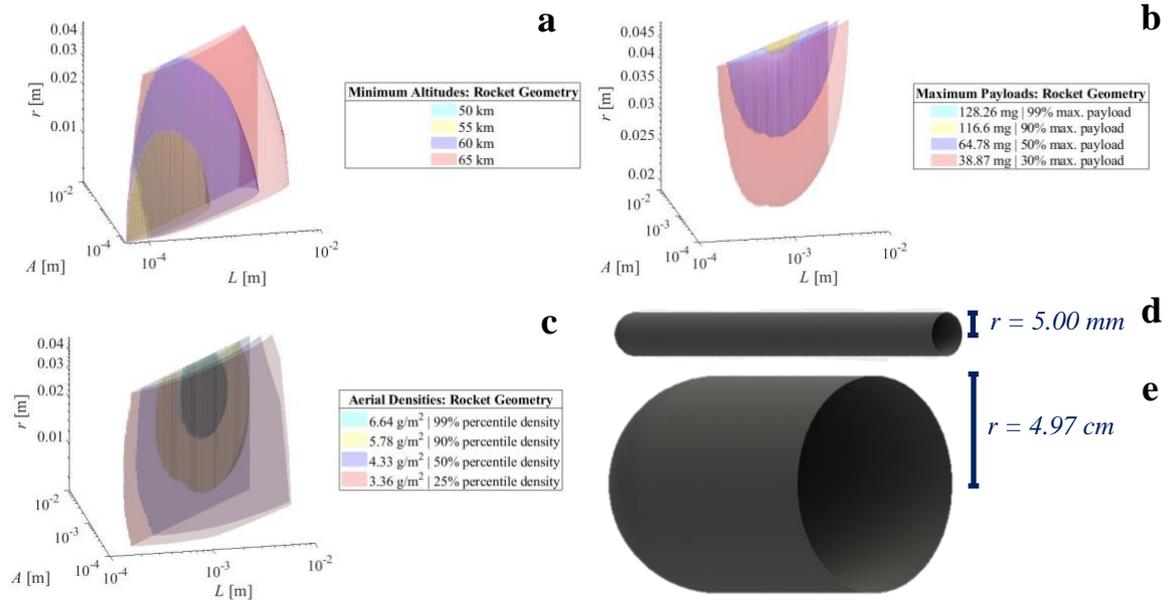

*Figure S14*: Minimum Altitude (a), Maximum Payload (b) and Areal Density (c) plots for the **D = 10 cm Rocket Geometry**. Here, the geometry that was able to levitate payload at minimum altitude (0.01 mg at 50 km) is shown in (d), while that which was able to levitate the maximum payload (129.56 mg at 80 km) is shown in (e).

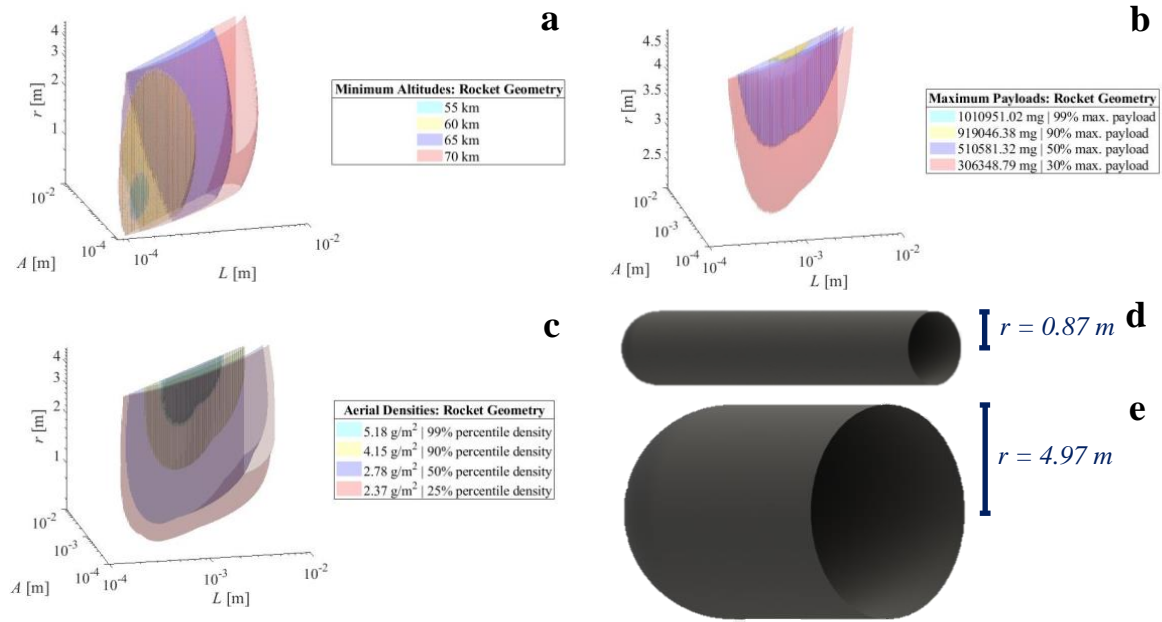

*Figure S15*: Minimum Altitude (a), Maximum Payload (b) and Areal Density (c) plots for the **D = 10 m Rocket Geometry**. Here, the geometry that was able to levitate payload at minimum altitude (2 132.57 mg at 55 km) is shown in (d), while that which was able to levitate the maximum payload (1 021 162 mg at 80 km) is shown in (e).

| Comparison of D = 10 cm and D = 10 m Rocket Geometries ||||||||
| Case || A | L | r | Surface Area (m²) | Areas Ratio | Payload (mg) | Altitude (km) |
| || Discretization of 500 points ||| | | | |
| --- | --- | --- | --- | --- | --- | --- | --- | --- |
| D = 10 cm | Min. Altitude | 0.11 mm | 0.12 mm | 0.50 cm | 0.001 | >100 | 0.01 | 50 |
| | Max. Payload | 1.24 mm | 1.25 mm | 4.97 cm | 0.05 | 6.02 | 129.56 | 80 |
| D = 10 m | Min. Altitude | 0.15 mm | 0.16 mm | 0.87 m | 59.39 | 24.98 | 2132.57 | 55 |
| | Max. Payload | 1.24 mm | 1.25 mm | 4.97 m | 467.23 | 6.02 | 1021162 | 80 |

*Table 13*: Combinations of A, L and r that returned the optimal rocket geometries described in *Figure S14* and *Figure S15* above.

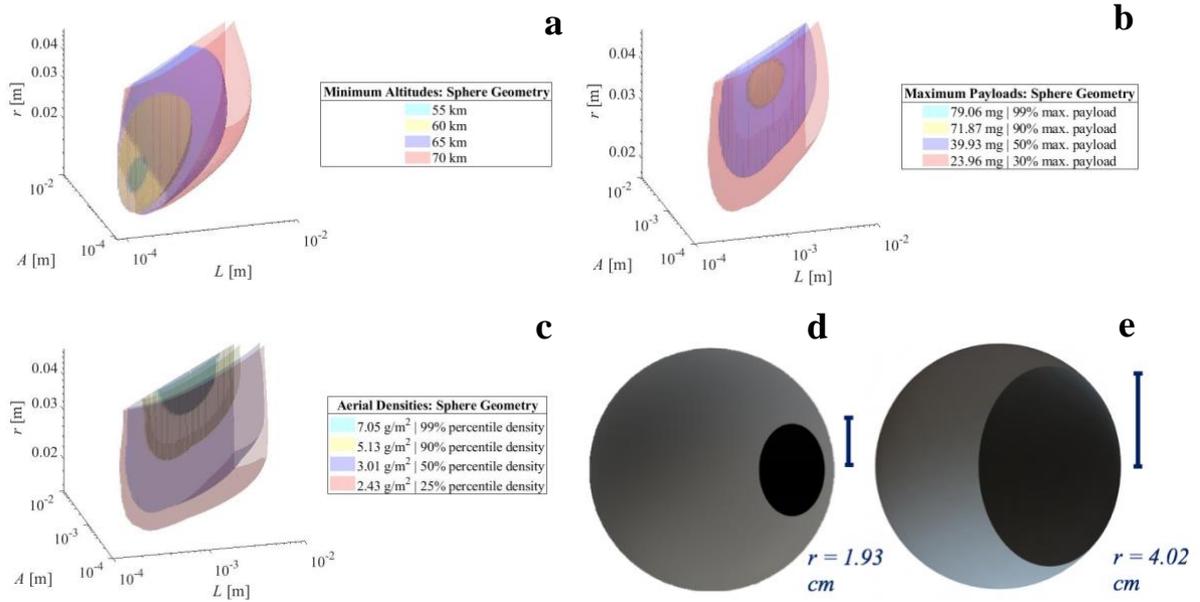

*Figure S16*: Minimum Altitude (a), Maximum Payload (b) and Areal Density (c) plots for the **D = 10 cm Sphere Geometry**. Here, the geometry that was able to levitate payload at minimum altitude (1.41 mg at 55 km) is shown in (d), while that which was able to levitate the maximum payload (79.86 mg at 80 km) is shown in (e).

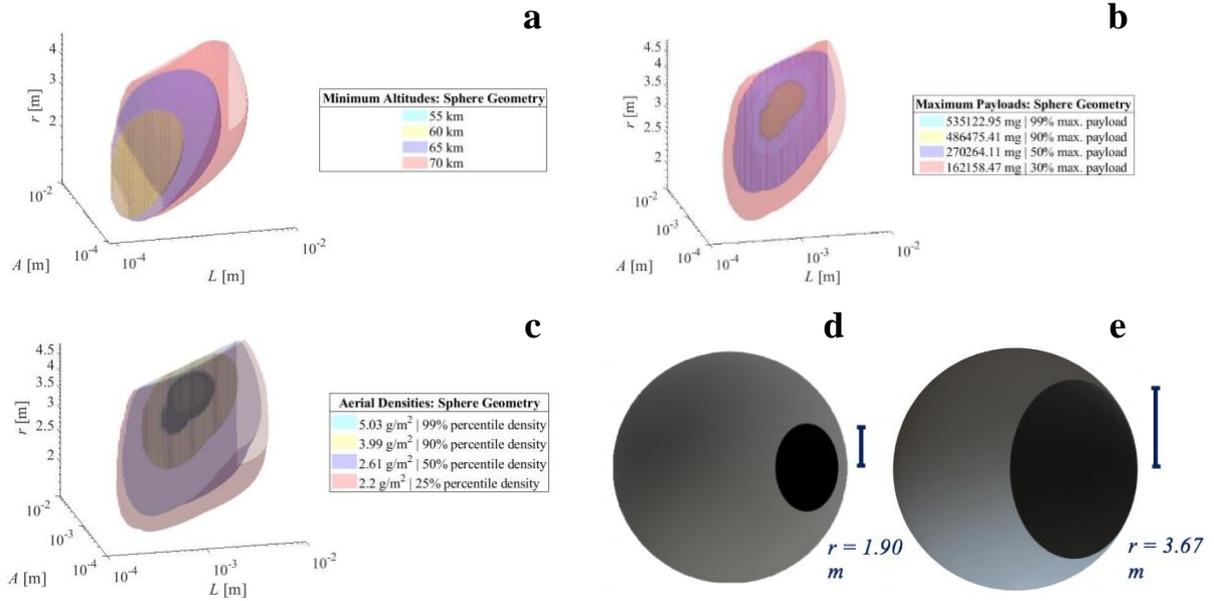

*Figure S17*: Minimum Altitude (a), Maximum Payload (b) and Areal Density (c) plots for the **D = 10 m Sphere Geometry**. Here, the geometry that was able to levitate payload at minimum altitude (831.92 mg at 55 km) is shown in (d), while that which was able to levitate the maximum payload (540 528 mg at 80 km) is shown in (e).

| Comparison of D = 10 cm and D = 10 m Sphere Geometries | | | | | | | | |
|---|---|---|---|---|---|---|---|---|
| Case | | A | L | r | Surface Area (m²) | Areas Ratio | Payload (mg) | Altitude (km) |
| | | Discretization of 500 points | | | | | | |
| D = 10 cm | Min. Altitude | 0.15 mm | 0.16 mm | 1.93 cm | 0.03 | 25.81 | 1.41 | 55 |
| | Max. Payload | 1.03 mm | 1.04 mm | 4.02 cm | 0.03 | 4.93 | 79.86 | 80 |
| D = 10 m | Min. Altitude | 0.15 mm | 0.16 mm | 1.90 m | 302.22 | 26.66 | 831.92 | 55 |
| | Max. Payload | 1.24 mm | 1.25 mm | 3.67 m | 263.63 | 6.23 | 540 528 | 80 |

*Table 14*: Combinations of A, L and r that returned the optimal sphere geometries described in *Figure S16* and *Figure S17* above.

# 3. Buckling Simulations in COMSOL

When creating high aspect ratio structures like the photophoretic aircraft, where the diameter-to-thickness ratio can reach $10^4$, buckling or other forms of structural failure may occur. However, nanocardboard is the best material to prevent or survive these effects because 1) Knudsen pumping action creates an outward pressure to help maintain the shape of the hollow 3D structure and 2) nanocardboard has ultra-high bending stiffness and can recover from sharp bending [R5].

To test the structures for buckling under their own weight, we performed COMSOL simulations for a few example geometries. We chose a 10-meter diameter sphere with a 4-meter diameter outlet and a 1 mm channel height, which is representative of structures optimized for maximum payloads. We also simulated a rocket geometry with a diameter of 4 m and a length of 10 m, and a cone of 10 m diameter, length of 8 m and outlet diameter of 6 m.

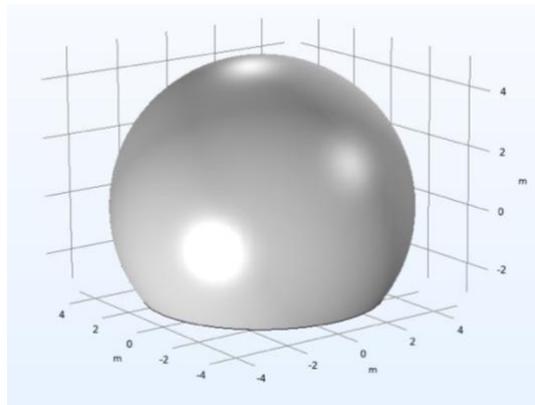

*Figure S18: Uniform sphere geometry within COMSOL. Similar models were used for the cone and rocket.*

In the simulations, we used an effective thin shell material whose weight, and bending and tensile stiffness correspond to those of nanocardboard. According to [R5], the tensile stiffness of nanocardboard is 3 times lower than for a uniform alumina shell of the same areal density. In contrast, the bending stiffness is increased by a factor ranging from 100 to 10000 for nanocardboard depending on the height of the channels and thickness of the shell; we chose the upper bound demonstrated in experiments even though the optimized structures are even taller than the 50 microns used in [R5] and should be even stiffer. We then use this enhancement to calculate the effective thickness, Young's modulus and density of the uniform shell that has the same stiffness properties as nanocardboard with face sheet thickness of 50 nm and 1 mm channel height, which represents the maximized payload case from the main text.

| *Channel height* | *Effective Thickness* | *Effective Young's Modulus* | *Effective Density* |
|---|---|---|---|
| 1 mm | 8.66 microns | 0.58 GPa | 115.5 kg/m^3 |

**Table 15**: *Corresponding effective properties due to nanocardboard TSEF and BSEF.*

We recorded the critical factor with the given properties in the COMSOL Shell simulations. A buckling factor greater than 1 means that the gravity force is insufficient to produce buckling of the shell under its own weight. The boundary conditions and loading included a simply supported edge for the outlet nozzle circle and gravity throughout the global coordinate system, representing the shell resting on the ground. We chose the "Extra Fine" physics-controlled meshing for all simulations. Geometric nonlinearities were included, and the results are shown in Figure S19.

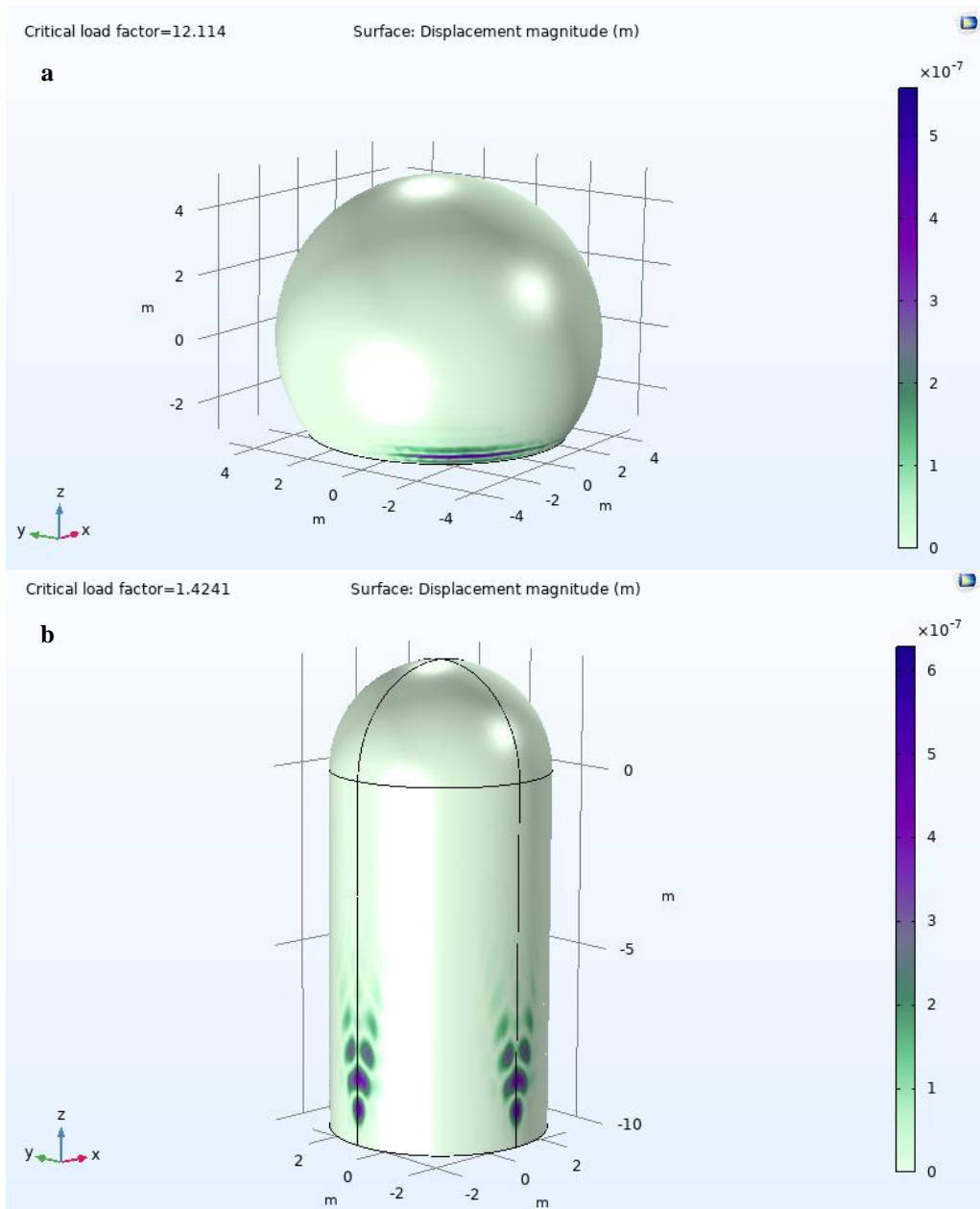

*Figure S19*: *(a) Uniform sphere example, with corresponding effective properties of nanocardboard, displacement magnitude plot with critical loading factor in the top left, (b) uniform rocket. Areas of maximum displacement represent where buckling will occur if load increases.*

The minimum critical loading factor for a uniform sphere with properties representing nanocardboard was **12.1** which means the weight to cause buckling would need to be more than 10 times the weight of the structure under given conditions. Another way of saying it is that the structure would not buckle until gravity or acceleration exceeded 12 g. The solver did not converge to a solution for the other geometries at standard gravity, suggesting buckling or lack of convergent buckling modes at high deformation, so we adjusted gravity down until a critical loading factor greater than 1 resulted. The cylindrical geometry is about 10 times too weak to support its weight, i.e., can only support its weight under 0.1 g gravity. In contrast, convergence was never achieved for the conical geometry even at gravity less than 1% of the of the standard value, indicating this geometry needs the most additional support, likely because the circular suspended top deforms the most of all geometries we considered. The results are shown in Table 15. Therefore, for the idealized cases in the paper, we assume the structure will successfully maintain shape and hold its own weight for the spherical case. but the other geometries require additional support structures.

| Geometry | Uniform Sphere, D = 10 m, D_out = 4 m | Rocket D = 4 m, L = 10 m | Cone D = 10 m, D_out = 6, L = 8 m |
|---|---|---|---|
| Gravitational load | 12.1 g | 0.1 g | N/A |

*Table 15*: *Corresponding gravitational load required for buckling with nanocardboard properties due to TSEF and BSEF.*

Should buckling occur due to additional imperfections not present in the simulations, there are additional ways to support the structures. For example, thin carbon fiber tubes or trusses can be used to support our proposed structure along the axis. Carbon fiber prepregs are available with thicknesses as low as 15 microns and tubes made from such thin carbon fiber composites would weigh less than the structure itself while preventing the relevant buckling modes. Frame-like structures can span the vertical diameter or length of the structure depending on the geometry case to further improve buckling resistance in areas of high deformation.